 \documentclass[12pt,preprint]{aastex}

\newcommand{\mum}{$\,\mu$m}
\newcommand{\Spitzer}{{\it Spitzer}}

\slugcomment{Accepted to PASP, July 19, 2007}

\begin{document}

\shortauthors{Stansberry et al.}
\shorttitle{MIPS 160\micron\ Calibration}

\title{Absolute Calibration and Characterization of the Multiband Imaging 
Photometer for Spitzer. III. \\
An Asteroid-based Calibration of MIPS at 160\mum}

\author{J. A. Stansberry\altaffilmark{1},
   K.D. Gordon\altaffilmark{1},
   B. Bhattacharya\altaffilmark{2},
   C.W. Engelbracht\altaffilmark{1},
   G.H. Rieke\altaffilmark{1},
   F.R. Marleau\altaffilmark{2},
   D. Fadda\altaffilmark{2}, 
   D.T. Frayer\altaffilmark{2},
   A. Noriega-Crespo\altaffilmark{2}, 
   S. Wachter\altaffilmark{2},
   E.T. Young\altaffilmark{1},
   T.G. M\"uller\altaffilmark{3},
   D.M. Kelly\altaffilmark{1},
   M. Blaylock\altaffilmark{1},
   D. Henderson\altaffilmark{2},
   G. Neugebauer\altaffilmark{1}
   J.W. Beeman\altaffilmark{4},
   E.E. Haller\altaffilmark{4,5}
   }
\altaffiltext{1}{Steward Observatory, University of Arizona, Tucson, AZ 85721}
\altaffiltext{2}{Spitzer Science Center, 220-6, Caltech, Pasadena, CA 91125}
\altaffiltext{3}{Max Planck Institute, D-85748 Garching, Germany}
\altaffiltext{4}{Materials Science Division, Lawrence Berkeley National Lab, 
   Berkeley, CA 94720}
\altaffiltext{5}{Department of Materials Science and Engineering,
   University of California at Berkeley, Berkeley, CA 94720}

\begin{abstract}

We describe the absolute calibration of the Multiband Imaging Photometer
for Spitzer (MIPS) 160\mum\ channel. After the on-orbit discovery of a
near-IR ghost image that dominates the signal for sources hotter than
about 2000~K, we adopted a strategy utilizing asteroids to transfer the
absolute calibrations of the MIPS 24 and 70\mum\ channels to the 160\mum\
channel. Near-simultaneous observations at all three wavelengths are
taken, and photometry at the two shorter wavelengths is fit using the
Standard Thermal Model. The 160\mum\ flux density is predicted from
those fits and compared with the observed 160\mum\ signal to derive the
conversion from instrumental units to surface brightness. The calibration
factor we derive is 41.7~MJy/sr/MIPS160 (MIPS160 being the instrumental
units). The scatter in the individual measurements of the calibration
factor, as well as an assesment of the external uncertainties inherent
in the calibration, lead us to adopt an uncertainty of 5.0~MJy/sr/MIPS160
(12\%) for the absolute uncertainty on the 160\mum\ flux density of a
particular source as determined from a single measurement.  For sources
brighter than about 2~Jy, non-linearity in the response of the 160\mum\
detectors produces an under-estimate of the flux density: for objects as
bright as 4~Jy, measured flux densities are likely to be $\simeq20$\%
too low. This calibration has been checked against that of ISO (using
ULIRGS) and IRAS (using IRAS-derived diameters), and is consistent with
those at the 5\% level.

\end{abstract}


\section{Introduction}

The Multiband Imaging Photometer for Spitzer (MIPS; Rieke et al. 2004)
is the far-infrared imager on the Spitzer Space Telescope (\Spitzer,
Werner et al. 2004). MIPS has three photometric channels, at 24, 70,
and 160\mum.  Like the other Spitzer instruments, the primary flux
density calibrators at 24 and 70\mum\ are stars. (IRAC: Reach et al.
2005; Fazio et al., 2004; Hora et al., 2004; and IRS: Houck et al.,
2004). The calibration for the MIPS 24 and 70\mum\ channels are presented
in companion papers by Rieke et al. (2007), Engelbracht et al. (2007:
24\mum) and Gordon et al. (2007: 70\mum). Here we present the calibration
of the 160\mum\ channel, and describe some unexpected challenges that
had to be overcome in performing the calibration.  The emission from
astonomical targets at this long wavelength is particularly useful in
characterizing the abundance of cold dust, which frequently dominates
the total emission from galaxies (e.g. Gordon et al., 2006; Dale et al.,
2005). The MIPS 160\mum\ channel has also contributed new insight into
the sources responsible for the previously unresolved cosmic infrared
background (Dole et al. 2006).

Very few calibrations exist in the 100--200\mum\ wavelength regime.
The Infrared Astronomical Satellite (IRAS; Neugebauer, 1984; Beichmann
et al., 1985) 100\mum\ channel, the 60 -- 200\mum\ channels of the
ISO Imaging Photopolarimeter (ISOPHOT, Schulz et al. 2002) aboard the
Infrared Space Observatory (ISO), and the Diffuse Infrared Background
Explorer (DIRBE, at 60 to 240 \mum; Hauser et al., 1998) aboard the
Cosmic Infrared Background Explorer (COBE, e.g. Fixsen et al. 1997)
relied on observations of solar system targets for their absolute
calibrations. The Far Infrared Absolute Spectrophotometer (FIRAS) on
COBE relied on observations of an external calibration target (Mather et
al. 1999).  In the case of IRAS, the calibration relied on observations
of asteroids to extrapolate the calibration of the 60\mum\ channel to
100\mum. In the case of ISOPHOT, a few asteroids were studied in great
detail, and their emission used as the basis of the absolute calibration
(M\"uller and Lagerros, 1998; 2002).  The primary reason these previous
missions relied on observations of asteroids (and planets) to calibrate
their longest-wavelength channels was sensitivity: the instruments could
not detect enough stellar photospheres at adequate signal-to-noise
ratio (SNR) over a wide-enough range of flux densities to support a
calibration. In part that was because the instuments had large beams
that were not well sampled by their detectors, leading to high confusion
limits to their sensitivity.

The original intention was to calibrate the MIPS 160\mum\ channel
using observations and photospheric models of stars. Compared to the
earlier missions, the MIPS detectors and electronics are significantly
more sensitive.  Also, the MIPS pixel scale, 16\arcsec, fully samples
the 40\arcsec\ beam provided by \Spitzer, resulting in lower confusion
limits.  After launch, the stellar calibration stategy was found to be
unworkable because a bright, short-wavlength ghost image impinged on
the array at nearly the same location as the 160\mum\ image (see below).
The strategy we adopted was similar to that employed by IRAS: namely to
use observations of asteroids in all three MIPS channels to transfer the
calibration from the MIPS 24 and 70\mum\ channels to the 160\mum\ channel.

\section{The Near-IR Ghost Image Problem}

Initial 160\mum\ commissioning observations of stars seemed to indicate
that the array was 10--15 times more responsive than expected from
pre-launch models and instrument characterization tests. However,
observations of cold sources seemed to confirm the expected responsivity
of the array. Within 4 months of the launch of Spitzer, we concluded
that for targets with stellar near-IR:160\mum\ colors, near-IR photons 
(with wavelengths $\simeq 1.6$\mum) were forming a ghost image on the
160\mum\ array.

The Ge detectors are sensitive to near-IR light because of their intrinsic
photoconductive response.  The desired response to 160\mum\ light, on the
other hand, arises from the extrinsic photoconductive response (achieved
by doping with Ga) coupled with mechanical stress applied to the pixels
(which extends the response from the normal 100\mum\ cutoff to about
200\mum).  Optical modeling eventually indicated that near-IR photons
diffusely reflected off the surface of the 160\mum\ short-wavelength
blocking filter were responsible for the ghost image.  That filter lies
near an intermediate focus in the optical train, and the reflected
photons form a poorly-focused ghost image on the array. By design,
the blocking filter is tilted relative to the light path to prevent
specularly reflected near-IR light from impinging on the array. However,
roughness on the surface of the blocking filter contributes a diffuse
component to the reflected near-IR light, and it is this diffusely
reflected light that forms the ghost image.

The near-IR light reflected from the blocking filter passes through the
160\mum\ bandpass filter (which has transmission in the near-IR of about
$10^{-3}$), but does not pass through the blocking filter. As a result,
the ghost image is quite bright in spite of the diffuse nature of the
reflection, having an intensity 10--15 times greater than the intensity
of the 160\mum\ image for sources with stellar colors.  The fact that
the ghost image nearly coincides with the image of 160\mum\ light on the
array (see Figure~1) made it difficult to identify the problem in the
first place, and also makes it very difficult to calibrate the relative
strengths of the two images.  Their relative strengths also depend on
the temperature of the source.  For a blackbody source spectrum (and
assuming that the effective wavlength of the ghost image is 1.6\mum),
objects with temperatures $\ge 2000$~K will suffer from a ghost image
comparable to or greater in brightness than the 160\mum\ image. Several
attempts have been made to overcome these uncertainties and difficulties,
and to characterize and calibrate the ghost-image directly, but have
met with quite limited success.

\section{Revised Calibration Strategy}

Asteroids were chosen as the new calibrators because of their very red
near-IR to 160\mum\ color, their ubiquity, and their range of brightness.
For typical asteroids the brightness of the ghost image will be at least
2000 times fainter than the 160\mum\ image, and so will not measurably
affect any calibration based on observations of asteroids.  Unfortunately,
asteroids also have several qualities that detract from their attraction
as calibrators: their far-IR SEDs are difficult to predict (due to
temperature variations across and within the surface), are time-variable
(due to rotation and changing distance from the Sun and observer),
and are poorly characterized at far-IR wavelengths. L and T dwarfs can
not be used because they are far too faint to be detected using MIPS
at 160\mum.

Because of the difficulty in predicting the 160\mum\ flux density
from a given asteroid for a particular observing circumstance,
we adopted a calibration strategy that relies on near-simultaneous
observations of asteroids at 24, 70 and 160\mum, and then bootstraps
the 160\mum\ calibration from the well-understood calibrations at 24 and
70\mum. Additionally, we have observed many asteroids, so that
we can use the average properties of the data to derive the calibration,
rather than relying on detailed efforts to model the thermal emission
of individual asteroids. The emission from asteroids at wavelengths
beyond 60\mum\ has only been characterized for a few objects (e.g. M\"uller 
and Lagerros, 1998; 2002), but those objects are all far too bright to 
observe with MIPS. 

\subsection{Faint \& Bright Samples}

Because the far-IR SEDs of asteroids are not well studied, we felt
that it was very important to characterize the thermal emission of
our calibration targets at both 24 and 70\mum\ to predict their
emission at 160\mum.  However, saturation limits introduce a complication
in trying to observe any particular asteroid in all 3 MIPS channels.
For a typical asteroid, the ratio of the flux densities, 24:70:160\mum,
is about 10:3:0.8.  The 24\mum\ channel saturates at 4.1~Jy in 1 second,
and somewhat brighter sources can be observed using the first-difference
image, which has an exposure time of 0.5 seconds. This limits the maximum
160\mum\ brightness that can be related back to well-calibrated 24\mum\
observations to about 0.5~Jy.  Sensitivity and confusion limits at
160\mum\ require that we observe asteroids brighter than about 0.1~Jy
at 160\mum. Thus, the dynamic range of the 160\mum\ fluxes that can
be directly tied to 24\mum\ observations is only a factor of 5, from
100~mJy to 500~mJy. The hard saturation limit at 70\mum, 23~Jy, does
not place any restriction on sources that can be observed at both 70
and 160\mum\ (the 160\mum\ saturation limit, 3~Jy, is about 1/2 of the
160\mum\ flux density from an asteroid with a 23~Jy 70\mum\ brightness).
These saturation-related restrictions lead us to adopt a 2-tiered
observation and calibration strategy.

\noindent {\it Faint Asteroids: 24\mum\ sample.} We observe asteroids
predicted to be fainter than $\sim 4$~Jy at 24\mum\ in all three MIPS
channels. The data are taken nearly simultaneously (typically less
than 30 minutes to observe all 3 channels, with nearly all of that
time being devoted to taking the 160\mum\ data). The short duration
of the observations limits potential brightness variations due to rotation of
the target (in addition, the targets were selected on the basis
of not exhibiting strong visible lightcurve variations). We then use the
observed flux densities at 24 and 70\mum\ to predict the flux density
at 160\mum\ using a thermal model (see below).  We also compute the
ratio of the {\it measured} 70\mum\ flux density to the 160\mum\ {\it
model prediction}, and use that ratio later to predict the 160\mum\
flux density for asteroids too bright to observe at 24\mum.

\noindent {\it Bright Asteroids: 70\mum\ sample.} For asteroids predicted
to be brighter than $\sim 4$~Jy at 24\mum, we observe only at 70 and 160\mum. 
We then use the average 70:160 color from the faint sample to predict the 
160\mum\ flux density from the 70\mum\ observation. This sample extends
the available dynamic range of the 160\mum\ observations by more than
a factor of two relative to the 24\mum\ sample alone, allowing us to
both measure the calibration factor up to the 160\mum\ saturation limit,
and to determine whether the response is linear.

\subsection{Limitations}

This strategy is subject to some limitations in addition to uncertainties
inherent to all absolute calibration schemes. The calibration we derive at
160\mum\ is wholly dependent on the MIPS calibrations at 24 and 70\mum,
and its accuracy can not exceed the accuracy of the calibration of
those channels. As described in Engelbracht et al. (2007), the absolute
calibration at 24\mum\ is good to 2\%; Gordon et al. (2007) show that
the 70\mum\ absolute calibration is good to 5.0\%.  These absolute calibration
uncertainties in the shorter channels translate into a 7\% uncertainty
on the predicted 160\mum\ flux density of any object with a a 24:70\mum\
color temperature of around 250~K (as our targets do).  This represents
the ultimate theoretical accuracy of the 160\mum\ calibration we can
derive via the methods described here.

As mentioned above, the dynamic range of the 160\mum\ fluxes that we can
relate to objects observed at both 24 and 70\mum\ is quite small. Thus,
the bright sample is critical for extending the dynamic range of the
calibration.  However, our predicted 160\mum\ fluxes rely on the average
70:160\mum\ model color of the faint sample, so the calibration is dependent
on the uncertainty in that color. The signal-to-noise ratio (SNR) of our
measurements at the shorter wavelengths is typically in excess of 50, so
their precision is not a major factor. However, the average 70:160\mum\
color we use depends on what we assume for the spectral emissivity
of asteroids.  There are hints in the ISO data that the emissivity of
some asteroids is depressed by $\simeq 10$\% in the far-IR (M\"uller \&
Lagerros, 2002), and model-based predictions that surface roughness may
also affect the slope of the far-IR thermal spectrum. Here we assume
that asteroids emit as gray-bodies, and use a thermal model that
does not incorporate the effect of surface roughness on the slope, and the
calibration we derive follows directly from those assumption. The full
impact of all of the uncertainties mentioned here on the accuracy of
the calibration are discussed in \S~8.1.

\section{Observations and Data Analysis}

\subsection{The Observations}

For each MIPS observing campaign, we used the JPL Solar System Dynamics
division's HORIZONs system (http://ssd.jpl.nasa.gov) to select main-belt
asteroids within the Spitzer operational pointing zone. From this set,
we selected objects with an albedo and diameter in the HORIZONs database
(primarily derived from the IRAS asteroid catalog, Tedesco et al. 2002).
For the purposes of observation planning only, we used the IRAS albedos 
and diameters to predict flux densities in the MIPS channels. We typically
selected a few to observe, picking those that could be observed in a
reasonable amount of time, would not saturate the detectors, and that did
not have significant lightcurve amplitudes (again, as indicated by the 
HORIZONs database).

102 individual observations of asteroids were made through the 28th
MIPS observing campaign (between December 2003 and January 2006). Of
those, 79 resulted in 160\mum\ detections with signal-to-noise ratios
$\ge 4$.  Thirty-three of those were 3-color (24, 70, and 160\mum)
observations of fainter asteroids, and 46 were 2-color (70 and 160\mum\
only) of brighter objects.  All observations were made using the MIPS
photometry Astronomical Observing Template (AOT), which provides dithered
images to improve point spread function (PSF) sampling and photometric
repeatability. The 160\mum\ array is quite small, having an (unfilled)
instantaneous field of view (FOV) of 0.8 by 5.3 arcminutes. The photometry
AOT, because of the dithers, results in a larger, but still restricted
2.1 by 6 arcminute filled FOV for the final mosaic. The diameter of the
first Airy minimum of the 160\mum\ PSF is 90\arcsec. After collecting
160\mum\ data using the standard dither pattern for a few observing
campaigns, we began taking those data by combining the AOT with small
map. This provided more sky around the target, and improved the sampling
of the PSF. Figure~1 shows a sample 160\mum\ image for a bright asteroid
resulting from such an observation.

\subsection{Data Analysis}
The data were analyzed using the MIPS instrument team data analysis tools
(DAT; Gordon et al., 2005). These tools have been used to develop the
reduction algorithms and calibration of the MIPS data, beginning during
ground-test, and continuing through on-orbit commisioning and routine
operations. The Spitzer Science Center data processing pipeline is
used to independently verify the algorithms and calibrations developed
through the instrument team DAT.  Both the SSC pipeline and the DAT
use the same calibration files (e.g. darks, illumination corrections),
and the same absolute calibration factors. Comparison of 160\mum\
photometry for data processed through the DAT and the SSC pipeline show
that the two agree to better than 1\%.  Data at 24 and 70\mum\ were
reduced, and photometry extracted, in exactly the same manner as all
other calibration data for those channels (see Engelbracht et al. 2007,
and Gordon et al. 2007). Because the exposure times at 24 and 70\mum\
were so short, the motion of the asteroids during those observations
was insignificant relative to the beam size in all cases. At 160\mum\
the beam is typically much larger than target motion, even though the integration
times in that channel were sometimes quite long. In the few instances
where object motion during the 160\mum\ observation was significant
(160\mum\ AOR execution times approaching one hour), we generated mosaics 
in the co-moving frame.

The basic processing of the 160\mum\ data is described in Gordon et al.
(2005). Briefly, each observation consists of multiple, dithered images.
During acquisition of each image, termed a data collection event (DCE),
the signal from the pixels is non-destructively sampled every 1/8 second.
The pixels were reset every 40th sample.  Cosmic rays are identified as
discontinuities in the data ramps, and slopes are then fit to the cleaned
ramps. Because the responsivity of the Ge:Ga array varies with time
and flux-history, internal relative calibration sources (stimulators)
are flashed every 8th DCE during data collection. Each slope image is
then ratioed to an (interpolated and background-subtracted) stimulator
image, and the result corrected for the measured illumination pattern
of the stimulators to produce a responsivity normalized image for each
dither position in an observation. Those images are mosaicked using
world coordinate system information to produce a final image of the
sky and target. The mosaics used in this analysis were constructed
using pixels 8\arcsec\ square, $\simeq$ 1/2 the native pixel scale of
the 160\mum\ array. This subsampling provides better PSF sampling and
aids in identifying outlier pixels during mosaicking.  Because the slope
image from each DCE is ratioed to a stimulator image, brightness in
the resulting mosaics is in dimensionless instrumental units which we
will refer to as ``MIPS160'' units, or simply MIPS160. The goal of the
calibration program is to derive the conversion (the ``calibration
factor,'' $CF$) between MIPS160 and surface brightness in units of, e.g.,
MJy/sr.

\section{Photometry and Aperture Corrections}

Figure~2 shows an azimuthally averaged radial profile of an observed
160\mum\ PSF, and compares it to model profiles generated using the
Spitzer PSF software (STinyTim, v1.3; Krist, 2002).  The measured profile
is derived from the observation of the bright (2.3~Jy) asteroid Papagena
(see Figure~1); other observations result in very similar PSFs.  Model
PSFs were generated assuming a source with a 250~K blackbody spectrum,
consistent with the temperatures we find for our sample. The models were
also generated using 5-times oversampling, resulting in model pixels
3\farcs2 square. As is seen for the other two MIPS channels (see
Engelbracht et al., 2007, and Gordon et al.  2007), the primary
difference between the model and observed PSFs is in the region of the
first Airy minimum.  However, suitably smoothed, the model PSF represents 
the observed PSF quite well. This is reflected in Figure~1, where
the overall morphology of the observed and model PSFs can be compared.
Figure~2 compares the radial profiles for the observed and model PSFs,
and shows the good agreement between the two.  The best-fit model PSF 
is smoothed using a boxcar with a width of 25\farcs6, corresponding to 
a width of 1.6 native pixels.

Because of the restricted FOV of the 160\mum\ images, we are forced to
use small apertures for performing photometry (this is in contrast to
the large apertures used to derive the 24 and 70\mum\ calibrations). Thus
the calibration at 160\mum\ depends more strongly on the aperture 
corrections.  We computed aperture corrections based on the model PSF shown
in Figures 1 and 2. The models offer two advantages over the observed
PSF: they are noiseless, and there is no uncertainty associated with
determining the background (particularly difficult at 160\mum\ because
of the restricted FOV).  The total flux in STinyTim model PSFs depends
on the model FOV: we utilized models 128\arcmin\ across in order to
capture most of the flux in the far-field of the PSF. We have extrapolated
the PSF to 512\arcmin\ using an Airy function, and integrated over
that much larger model to constrain the magnitude of any bias in our
aperture corrections stemming from their finite FOV. Those calculations
indicate that only 0.1\% of the flux from a source falls in the region
between 128\arcmin\ and 512\arcmin: we conclude that our aperture 
corrections are not significantly biased by our use of the 128\arcmin\
models. Later we show that our calibration, when applied to extended
sources, gives results consistent with ISO to within 6\%. That
agreement provides some additional confidence in the accuracy of our 
aperture corrections.

Application of the model-based aperture corrections to observed PSFs
revealed that for apertures $\le 48\arcsec$ in radius the measured
flux depended on aperture size. The reason is the small but systematic
difference between the observed and model PSFs at radii of $\simeq
10\arcsec$--20\arcsec, which can be seen in Figure~2. To correct this,
we have adopted a hybrid approach to computing the aperture corrections,
using the smoothed model PSFs for apertures with radii $\ge 48\arcsec$,
and observed PSFs for smaller apertures. We used observations of 9
asteroids observed using a small 160\mum\ map (giving a somewhat
larger FOV, as noted earlier), and with fluxes near 1~Jy for the
computation. (We also compared these asteroid-based corrections to
those based on Pluto (with a color temperature of 55--60~K), and found no
measurable difference).  The empirical corrections are normalized to the
model correction for the 48\arcsec\ aperture.  Table~1 lists the resulting
hybrid aperture corrections for a selection of photometric aperture sizes,
with and without sky annuli, and for a range of source temperatures.
Note that these corrections can only accurately be used for sources that
are relatively cold (significantly less than 2000~K) -- otherwise the
near-IR ghost image both alters the PSF, and becomes comparable to or
brighter than the 160\mum\ image. We have verified that the corrections
in Table~1 result in photometry that is independent of aperture size by
analyzing 29 cluster-mode asteroid observations, where the targets ranged
in brightness from 0.1--4~Jy. The variation with aperture size shows no
monotonic trend, and the results for all apertures agree to within 1\%.

We performed photometry on our 160\mum\ images using an aperture
24\arcsec\ in radius.  The small aperture allowed us to increase
the SNR of our photometry for the faintest asteroids, and thereby to 
extend the calibration to somewhat fainter flux densities than would
have been possible otherwise. The aperture photometry was corrected to
total counts using the aperture correction in Table~1.  Photometry at 24
and 70\mum\ was performed exactly as it was to derive the calibrations
in those channels, and as described in Engelbracht et al. (2007) and
Gordon et al. (2007). Because a number of our brightest asteroids were
in the non-linear response regime at 70\mum ({\it i.e.} above a few Jy), 
we have used PSF-fitting (using the StarFinder package: Diolaiti et al.
2006) to do all of the 70\mum\ photometry used here. We attempted to
analyze the 160\mum\ data using PSF-fitting as well, but the resulting
photometry displayed more scatter than did the aperture photometry.
We believe this was due to the restricted FOV of the mosaics, and the 
presence of spatial structure (artifacts) in the images, particularly 
for fainter sources. An area of concentration in the future will be 
implementing more robust PSF-fitting algorithms for use at 160\mum.

\section{Color Corrections}

The effective wavelengths of the MIPS channels, defined as the average
wavelength weighted by the spectral response function, $R(\lambda)$,
are $\lambda_0=$ 23.68, 71.42 and 155.9\mum.  The color corrections,
which correct the observed in-band flux to a monochromatic flux density
at the effective wavelength, are defined by:
$$
K ={{{1\over{F(\lambda_0)}}\int F(\lambda) R(\lambda) d\lambda } \over
    {{1\over{G(\lambda_0)}}\int G(\lambda) R(\lambda) d\lambda } }.
$$
Here $F(\lambda)$ is the spectrum of the source, $G(\lambda)$ is the
reference spectrum, $\lambda$ is wavelength, $F$ and $G$ are in units
of photons/sec/cm$^2$/$\mu$m, and $R$ is in units of $e^-$/photon. As
defined here, the observed flux should be divided by $K$ to compute
the monochromatic flux density.  The MIPS response functions can be
obtained from the Spitzer web site (http://ssc.spitzer.caltech.edu/mips).
For MIPS, the reference spectrum, $G$, is chosen as a $10^4$~K blackbody. While
we refer to the 24, 70 and 160\mum\ channels, we have used the actual
effective wavelengths of those channels for all quantitative analyses.
For reference, the zero magnitude flux density at 155.9\mum\ is $160 \pm
2.45$~mJy. Because the asteroids are much colder (with typical 24:70\mum\
color temperatures around 250~K), we had to apply color corrections
to convert the measured fluxes to monochromatic flux densities at
the effective wavelengths. The color corrections for all three MIPS
channels and representative source spectra are given in Table~2. In all
three channels they are slowly varying functions of temperature above
temperatures of 100~K, and also deviate only a few percent from unity
at those temperatures. For objects with data at both 24 and 70\mum, the
color corrections were computed iteratively based on the 24 and 70\mum\
flux densities.  For the brighter targets lacking 24\mum\ data, we assumed
a temperature of 251~K (see Figure~4), and applied the corresponding
color correction.

\section{Thermal Modeling}

The Standard Thermal Model (STM, Lebofsky \& Spencer 1989) is the most
widely used (therefore ``Standard'') model for interpreting observations
of thermal emission from small bodies in the asteroid main belt and the
outer Solar System ({\it c.f.} Campins et al.  1994; Tedesco et al. 2002;
Fern\'{a}ndez et al. 2002; Stansberry et al. 2006).  The model assumes
a spherical body whose surface is in instantaneous equilibrium with the
insolation, equivalent to assuming either a thermal inertia of zero,
a non-rotating body, or a rotating body illuminated and viewed pole-on.
In the STM the subsolar point temperature is
\begin{equation}
T_0 = [S_0(1-p_Vq)/(\eta\epsilon\sigma)]^{1/4}\;,
\end{equation}
where $S_0$ is the solar constant at the distance of the body, $p_V$ is
the geometric albedo, $q$ is the phase integral (assumed here to be
0.39, equivalent to a scattering asymetry parameter, $G = 0.15$ (Lumme
and Bowell 1981; Bowell et al. 1989)), $\eta$ is the beaming parameter,
$\epsilon$ is the emissivity (which we set to 0.9), and $\sigma$ is the
Stefan-Boltzmann constant. Given $T_0$, the temperature as a function of
position on the surface is $T = T_0 \mu^{1/4},$ where $\mu$ is the cosine
of the insolation angle.  The nightside temperature is taken to be zero.
Surface roughness leads to localized variations in surface temperature
and non-isotropic thermal emission (beaming). When viewed at small phase
angles, rough surfaces appear warmer than smooth ones because the
emission is dominated by warmer depressions and sunward-facing slopes.
This effect is captured by the beaming parameter, $\eta$. Lebofsky et
al. (1986) found $\eta = 0.76$ for Ceres and Vesta; the nominal range for
$\eta$ is 0 to 1, with unity corresponding to a perfectly smooth surface
(Lebofsky \& Spencer 1989).

The purpose of our thermal modeling is to use the measured 24 and/or
70\mum\ flux densities to predict the 160\mum\ flux density for that
target.  First we correct the flux density from the observed phase angle
(typically about 20\degr\ for our targets) to 0\degr\ using a thermal
phase coefficient of 0.01~mag/\degr\ (e.g. Lebofsky et al. 1986).
We then use the absolute visual magnitude ($H_V$, defined for a phase
angle of 0\degr) from Horizons and the relation (e.g. Harris, 1998)
$D=1329\times10^{-H_V/5}\,p_V^{-1/2}$ to compute the target diameter
(where $D$ is the diameter in km, and $p_V$ is the visible geometric
albedo).  Target diameter and albedo are varied until a fit to the
observed flux density is achieved. For targets observed at both 24 and
70\mum, the beaming paramter is also varied in order to simultaneously fit
both MIPS bands and the visual magnitude. The fitted physical parameters
are then fed back into the STM to predict the 160\mum\ flux density.

Figure~3 illustrates the measured spectral energy distribution (SED)
for one of our targets. Also shown are a blackbody and STM fit to the
24 and 70\mum\ points. The blackbody and STM fits are indistinguishable
at the MIPS wavelengths, but small deviations can be seen on the short
wavelength side of the emission peak. For the purpose of calibrating
the 160\mum\ channel, we simply require a reliable way to predict the
160\mum\ flux density by extrapolation from the shorter wavelengths.
As the figure demonstrates, the details of the short-wavelength SED do
not appreciably affect the predicted 160\mum\ flux density. Indeed, we
have performed the calibration using both STM and blackbody predictions,
and the results are consistent with each other to within better than 1\%.

\section{Results} 

\subsection{The 24\mum\ Subsample}
Table~3 summarizes our measurements of targets in the 24\mum\ sample.
Aperture- and color-corrected flux densities are given for the 24 and
70\mum\ measurements.  The 160\mum\ data are given in the instrumental
units, MIPS160, described in \S~4.2.  As for the shorter wavelengths,
the 160\mum\ measurements have been aperture- and color-corrected.
The 24\mum\ sample makes up one half of the full data set, and covers
the faint end of the sample. These observations also allow us to directly
determine the color temperatures (used to compute color corrections for
individual observations within the sample, and to predict an average
color temperature, used to compute color corrections for the 70\mum\
sample). We also use the 24\mum\ sample to compute the average 70:160\mum\
model color for asteroids, which we use to predict 160\mum\ fluxes for
the 70\mum\ sample.

Figure~4 shows the color temperatures of the objects in the 24\mum\
sample, determined by fitting a blackbody to the photometry in those
channels. The temperatures are fairly tightly clustered, with an average
and standard deviation of $\simeq 251\pm 25.6$~K. The temperatures are
plotted vs. predicted 160\mum\ flux density. In the context of this
figure (only), the prediction is simply the extrapolation of the fitted
blackbody curve to 160\mum.  Although the range of predicted 160\mum\
flux densities for the 24\mum\ sample is only a factor of 5, there is no
apparent trend of color temperature.  Because the temperatures are fairly
similar amongst all the targets, the predicted 160\mum\ flux density is
to first order a measure of the overall apparent thermal brightness of
the targets. It then reflects a combination of the influences of distance
(helio- and Spitzer-centric), albedo, and size. It might be expected that
if any of these things were biasing our results, or imposing a systematic
trend in the predicted 160\mum\ flux density (e.g. if our brightest
targets were systematically hotter), it would be apparent in this figure.

Given the fairly narrow range of color temperatures we see for the objects
in the 24\mum\ sample, and the insensitivity of the model spectra from
24 to 160\mum\ to details of the thermal models, we expect the 70\mum\
to 160\mum\ color of the asteroids to be quite constant. Figure~5 shows
the ratio of the measured 70\mum\ flux density to the predicted 160\mum\
flux density for each asteroid in the 24\mum\ sample.  As expected, the
color is tightly clustered, with a mean value of 3.77, and a root-mean-square
(RMS) scatter of 0.095, or 2.5\%. Under the assumption that asteroids do not
posess any strong emissivity variations vs. wavelength in the far-IR,
we use this color ratio to interpret our data for the brighter asteroids.

\subsection{The 70\mum\ Subsample} 
Table~4 summarizes our measurements of targets in the 70\mum\ sample, and
is exactly like Table~3 except for the lack of 24\mum\ data.  Making use
of the average 70:160\mum\ color from the 24\mum\ sample, we compute
the predicted 160\mum\ flux density for the 70\mum\ sample. The
uncertainty on the 160\mum\ prediction is derived from the uncertainty
in the 70\mum\ measurement root-sum-square (RSS) combined with the 2.5\%
uncertainty in the average 70:160 color.

\section{Calibration Factor} 
Figure~6 shows the calibration factor ($CF$) we derive from our observations
of both the 24\mum\ and 70\mum\ samples, as a function of the predicted
160\mum\ flux density. The calibration factor is defined as the predicted
flux density at 160\mum\ divided by the (aperture- and color-corrected)
brightness in instrumental units (MIPS160), and by the area of a pixel
in steradians.

Of the 102 individual observations, 23 were rejected on the grounds of
having 160\mum\ SNR$<$4; three more were rejected for having a measured 
160\mum\ flux density more than twice the prediction (these were all for
very bright sources, and the discrepancy is due to poorly compensated
non-linear response in the 70\mum\ channel resulting in predictions that
were too low.  Figure~6 shows the remaining 76 values of the calibration
factor. There is a fairly clear trend of increasing calibration factor
for predicted flux densities greater than about 2~Jy. We attribute this
trend to a non-linear response of the detectors for bright targets. This
effect is similar in magnitude to that seen at 70\mum, also at flux
densities greater than about 1--2~Jy (Gordon et al. 2007). For the moment
we exclude the 19 points above 2~Jy from consideration. Taking the
points below 2~Jy, we compute the average and RMS scatter, and identify
as outliers 8 points that deviate from the mean by more than 1.5 times
that scatter (indicated by circled points in Figure~6). We use the
weighted mean of the remaining 49 values to compute the calibration
factor for the MIPS 160\mum\ channel.  Use of the weighted mean ensures
that a source with zero flux produces zero response if all of the inputs
to the calibration (e.g. dark current, linearity) are perfectly known.

The weighted mean calibration factor is $CF = 41.7$~MJy/sr/MIPS160, and
the RMS scatter is 4.82~MJy/sr/MIPS160. This suggests an uncertainty
of 11.6\% for the determination of the flux density of a particular
source based on a single meausurement.  The formal uncertainty
on the average calibration factor is 0.69~MJy/sr/MIPS160, or only
1.6\%, but this value clearly underestimates the uncertainty that
should be assumed when interpreting 160\mum\ photometry (see below).
The average calibration factor and RMS scatter are shown in Figure~6
as the horizontal dashed lines. Below we discuss other sources of
uncertainty in the calibration. The final value and uncertainty we adopt
are $41.7\pm5.0$~MJy/sr/MIPS160 (equivalent to a 12\% uncertaintiy).
This calibration is valid for sources with 155.9\mum\ flux densities
$\le2$~Jy.

We also computed a weighted linear fit to the data, but in this
case include those points with predicted 160\mum\ flux densities
$>2$~Jy.  Based on the linear fit, $CF = (39.24 + 2.58\times
P_{160})$~MJy/sr/MIPS160, where $P_{160}$ is the predicted 160\mum\
flux density. The formal uncertainties on the intercept and slope
from the linear fit are 1.29~MJy/sr/MIPS160 and 0.76~MJy/sr/MIPS160/Jy,
respectively, indicating that the slope is significant at the $3.4\sigma$
level. This reflects the influence of the response non-linearity above
2~Jy, and can be used to provide an approximate calibration of targets
with flux densities $>2$~Jy. Inspection of the points in Figure~6
suggest that the non-linearity may affect photometry at the 20\% level
for targets with flux densities near 4~Jy, somewhat more than would
be derived based on the linear fit to the data.

\subsection{Uncertainty on the 160\mum\ Absolute Calibration}

As suggested above, observers are typically more interested in the
uncertainty they should assume for the flux density they determine
from a single observation of a target than they are in the formal
uncertainty on the calibration factor determined from an ensemble. Here
we  compare the 11.6\% uncertainty estimated above to the uncertainty
we would expect given the other uncertainties in the inputs to
the calibration. The relevant uncertainties to consider are: 1) the
photometric repeatability at 160\mum, 2) the uncertainties in the 24\mum\
and 70\mum\ calibrations, 3) systematic uncertainties associated with
color and aperture corrections, and 4) uncertainties inherent to the
models used in the calibration.

We have assessed the photometric repeatability of the 160\mum\ channel
two ways. Because we have relatively few repeated observations of stable
(i.e. non-asteroidal), red sources, we analyzed 81 160\mum\ observations
of a stellar calibrator (HD 163588), and found that those measurments
exhibited an RMS scatter of 3.4\%. While those data are severely impacted
by the short-wavelength ghost, they do provide a valid measure of the
repeatability delivered by the readout electronics and the end-to-end
data analysis for a very bright source. We have also analyzed 5 160\mum\
observations of IRAS~03538-6432, which has a very red near-IR:160\mum\
color, and a 160\mum\ flux density of $\simeq1.04$~Jy (Klass et al. 2001),
finding an RMS scatter of 5.5\%. We adopt 5\% as our current estimate 
of the repeatability.

The uncertainties in the calibrations of the shorter MIPS bands
are estimated to be 2\% (24\mum: Engelbracht et al. 2007) and 5\%
(70\mum: Gordon et al. 2007). As noted earlier, taken in combination
and ignoring any other uncertainties, these place a lower limit on
the 160\mum\ calibration uncertainty of 7\%. The color corrections we
have applied are very modest (a few percent), and so are unlikely to
contribute significantly to the calibration uncertainty. The 24 and
70\mum\ photometry was done identically to the way it was done for the
calibrations of those bands, and so should not impose any additional
uncertainty or systematic bias on the results used here.

The 160\mum\ aperture correction we used, 2.60, is large and is probably
uncertain at the level of a few percent.  Uncertainty in the aperture
correction will be irrelevant if others use the same aperture ({\it i.e.}
24\arcsec, with a sky annulus of 64\arcsec--128\arcsec) and correction
to perform photometry of point sources, and we encourage observers to
use this aperture when practical. However, we can not assume that such
will be the case. Checks of 160\mum\ measurements of extended sources
(see below) against previous missions suggest agreement to within about
6\%, suggesting that our aperture corrections are reasonably accurate.
As noted earlier, we find no evidence to suggest that the aperture
correction for the 24\arcsec\ aperture is any more uncertain than that
for a 48\arcsec\ aperture, where the aperture correction is a more modest
(and model-based) 1.60.  For lack of good 160\mum\ observations to further
assess the uncertainty in the aperture corrections, and based on our
experience with the 24 and 70\mum\ calibrations, we adopt an uncertainty
of 3\% for our 160\mum\ aperture corrections.  This uncertainty should
be interpreted as applying to the 48\arcsec\ aperture, and as being
empirically verified as transferable to the 24\arcsec\ aperture.

The final uncertainty in the calibration is associated with the
assumptions inherent in the Standard Thermal Model, particularly the
spectral emissivity in the 24\mum\--160\mum\ range. As noted earlier, we
have assumed a gray emissivity, whereas there are suggestions from ISO
observations that the emissivity of some asteroids may decline by 10\%
or so in this region (e.g. M\"uller and Lagerros, 2002). We find that
our 24 and 70\mum\ measurements of asteroids, when fit independently
with the STM, give diameters for the targets that agree to within 3\%,
with an RMS scatter of 5\% (the 70\mum\ diameters being smaller). This
suggests that there is no strong decrease of emissivity for the asteroids
in our sample between 24 and 70\mum\ (because those calibrations are
derived solely from observations of stars). Unfortunately we can not
make a similar argument about emissivity in the range 70--160\mum\ based
on our data. We adopt an uncertainty of 5\% to account for our lack of
knowledge of the spectral emissivity at 160\mum, and as being consistent
with the lack of evidence for any measurable emissivity trend from 24--70\mum.

If we RSS combine the uncertainties just discussed, we predict that
the 160\mum\ calibration should be accurate to 10.4\%, which is very
consistent with the 11.6\% uncertainty estimated from the RMS scatter of
the calibration factor values in Figure~6. While the combined effect
of the calibration uncertainties at 24 and 70\mum\ are the largest
single contributor to the 160\mum\ uncertainty, the other uncertainties
together are at least as important. Given that emissivity effects would
result in a systematic bias in our calibration, we should not
really RSS it with the other uncertainties. If we RSS-combine the other
uncertainties, and then simply add the 5\% uncertainty for emissivity
effects, we predict a worst-case uncertainty of 14.1\% in the calibration
(worst-case because it assumes that the net effect of the random
uncertainties combine constructively with the emissivity uncertainty).
Given the general agreement in the magnitude of these estimates and that
based on the RMS scatter of the measurements of $CF$ itself, we
adopt an uncertainty of 12\% for the absolute calibration of the 160\mum\
channel of MIPS. 

\subsection{Calibration Cross Checks}

Soon after the launch of \Spitzer, observations of a few targets that
have well-studied SEDs in the 160\mum\ region were made, and formed
the basis of the initial calibration. These included observations
of a few asteroids (those data were included in the analysis above),
which led to $CF = 41.6\pm8.5$~MJy/sr/MIPS160. Observations of K-giant
calibration stars were affected by the near-IR ghost, but after roughly correcting for
the ghost, those data indicated $CF = 37.8\pm11.3$~MJy/sr/MIPS160. Early
science observations of Fomalhaut were also analyzed, and indicated $CF =
39.8\pm6.0$~MJy/sr/MIPS160.  We also analyzed early science data for M33
(Hinz et al., 2004), NGC 55, NGC 2346, and the Marano Strip, which,
taken together, indicated $CF = 46.8\pm12$~MJy/sr/MIPS160.  All of
these results lead us to adopt an initial calibration for the 160\mum\
channel of $CF = 42.5\pm8.5$~MJy/sr/MIPS160. Gordon et al. (2006) have
compared MIPS 160\mum\ measurements of M31 to DIRBE and ISO measurements,
finding excellent agreement. All of these provide a sanity check of
the new calibration, because it is only 1.9\% lower than the initial 
calibration.

More recently we have compared MIPS measurements of a few ULIRGs to
ISO measurements of the same objects, and to the IRAS results for the
asteroids observed for the MIPS 160\mum\ calibration program. In both
of these cases we have included comparisons at the shorter MIPS bands
as well as at 160\mum. The comparisons at the shorter wavelengths serve
two purposes. Because both the 24 and 70\mum\ calibrations are entirely
based on observations of stars, any short-wavelength spectral leaks
present in those channels would bias photometry of cold sources such
as ULIRGs and asteroids: the comparisons serve to confirm the lack of
such leaks. Because the 160\mum\ calibration is derived directly from
the shorter MIPS bands, the comparisons at those wavelengths also serve
to confirm the validity of the 160\mum\ calibration, even though it
(unlike for the shorter bands) is based on observations of red sources.

We reduced Spitzer archive data for the ULIRGs IRAS 03538-6432 (5
epochs), IRAS 13536+1836, IRAS 19254-7245 and IRAS 20046-0623 (1 epoch
for each), and measured their flux densities at 70 and 160\mum. The
70\mum\ flux densities for the first three was within a few percent
of the values we would expect based on the ISO photometry reported by
Klaas et al. (2001). In particular, for the first two, the MIPS and ISO
results agreed to better than a percent.  The 160\mum\ flux densities
were 5\% higher than expected from the ISO data on average. Again, for
IRAS 03538-6432 the agreement was within 1\%. The MIPS data for
IRAS 20046-0623 gave 70 and 160\mum\ flux densities 25\%--30\% lower
than would be expected from the ISO data, but there is no obvious
reason for this discrepancy (e.g. no bright background objects that
might have fallen within the ISO beam).

We have also fitted our 24 and 70\mum\ observations of asteroids with
the STM, deriving diameters for all our targets. The diameters we
derive by fitting the two bands independently (for the faint sample)
agree quite well: the mean and RMS scatter of the ratio of
the diameters determined at 24\mum\ to those determined at 70\mum\
being 1.02 and 0.051, respectively. This confirms that the calibrations
of these two bands are very consistent when applied to observations
of red sources. The small deviation of this ratio from unity has a
formal significance of $2.8\sigma$, but could easily be due to 
the failure of the simple assumptions of the STM to fully describe
the thermal emission. We also have compared the diameters determined
from our data to the diameters derived from IRAS data (the SIMPS
catalog, Tedesco et al. 2002). The average and RMS scatter of the
ratios of the MIPS diameters to the IRAS diameters at 24\mum\ are
1.01 and 0.09, while at 70\mum\ they are  0.99 and 0.10. We conclude
that our calibration in those bands is entirely consistent with
the IRAS calibration; by inference the 160\mum\ calibration should
also be consistent with IRAS.

\subsection{Extended Source Calibration}

We also checked the calibration on extended sources at 160\mum, using
observations of a handful of resolved galaxies which were observed
by ISOPHOT using the C\_160 broad band filter ($\lambda_{ref} =
170$~\micron).  The galaxies used for this comparison are M31 (Haas et
al. 1998; Gordon et al. 2006), M33 (Hippelein et al. 2003; Hinz et al.
2004), M101 (Stickel et al. 2004; Gordon et al. 2006, in prep.), and
NGC3198, NGC3938, NGC6946, and NGC7793 (Stickel et al. 2004; Dale et al.
2005, 2007). These objects range in diameter from 5--10\arcmin\ (the NGC
objects) to $\ge 0.5\degr$ (the Messier objects), and so are all highly
resolved by both MIPS at 160\mum (40\arcsec\ FWHM) and ISOPHOT at 170\mum\
(90\arcsec\ pixels). We applied color corrections to the MIPS and ISOPHOT
measurements, and corrected for the difference in wavelengths, assuming
the emission has a color temperature of 18~K. The resulting average ratio
and uncertainty in the mean of the MIPS 160\mum\ to ISOPHOT 170\mum\
flux densities is $0.94\pm0.06$. If the emissivity of the dust in these
galaxies is proportional to $\lambda^{-2}$, the expected ratio of the
measurements is 1.00, consistent to within the uncertainty in the measured
mean. Thus the MIPS and ISOPHOT extended-source calibrations near 160\mum\
are entirely consistent with one another. These comparisons also indicate
that the MIPS point-source derived calibration at 160\mum\ is directly
applicable to observations of extended sources, and by inference that
the aperture corrections in Table~1 are accurate to within a few percent.

\subsection{160\mum\ Enhanced AOT: Calibration and Sensitivity}

In Spring of 2007 a new 160\mum\ photometry observing template (the
``Enhanced AOT'' was made available. The goal of the new template is
to allow 160\mum\ photometry data to be time filtered, as has been
done all along for the 70\mum\ data. A limited number of observations
(3) taken using the enhanced 160\mum\ AOT were available at the time
of this writing.  In each case, the same target was observed using the
standard 160\mum\ AOT as well. 

All of these data were reduced in the standard manner, as described
earlier.  In addition, the enhanced AOT data were processed by applying
a high-pass time-domain filter to the time series for each pixel (this
filtering process is a standard part of the reduction at 70\mum: Gordon
et al. 2005; 2007).  Because a dither is performed between all images,
the filter preserves the signal from point sources while suppressing
elevated noise levels that result from signal drifts in un-filtered data
products.  Such filtering can not reliably be applied to data from the
standard AOT because the dithers never completely move the source out
of the FOV of the array. The result is that time-filtering erodes flux
from the target source, and does so in a way that is flux dependent.
The enhanced AOT implements a wider dither pattern, providing enough
data away from the source that the filter works well.

Photometry on the standard AOT, enhanced AOT without time-filtering,
and enhanced AOT with time-filtering was measured as described earlier.
We draw preliminary but encouraging conclusions based on these initial
results. 1) Photometry measured on the standard and enhanced AOT data
agree to within about 5\%, except on bright ($>1$~Jy) sources, where the
time-filtered product gives systematically lower fluxes (at about the
10\% level). Thus, the enhanced AOT should only be utilized for sources
expected to be fainter than about 1~Jy. 2) The time-filtered enhanced AOT
data provides significant sensitivity improvements over the standard AOT,
unfiltered data. We computed the $1\sigma$, 500 second noise-equivalent
flux density (NEFD, frequently referred to as ``sensitivity'').  For the
old AOT NEFD$= 35$~mJy, while for the enhanced AOT NEFD$=22$~mJy.
Thus the enhanced AOT improves the point-source sensitivity of the
160\mum\ channel by about 35\%.  We lacked sufficient data to compare the
repeatability of the enhanced AOT relative to the old AOT, but expect
that it may result in some significant gains, particularly for faint
sources and/or higher backgrounds.

\section{Summary}

We have undertaken a program to calibrate the MIPS 160\mum\ channel
using observations of asteroids. The strategy employed was statistical in
nature: rather than perform detailed modeling of a few asteroids to try
and accurately predict their 160\mum\ flux density for our observing
circumstances, we instead rely on the average emission properties
of asteroids in the spectral range 24 -- 160\mum\ to allow us to
transfer the calibration of our 24 and 70\mum\ channels to the 160\mum\
channel. Our 24 and 70\mum\ data from 51 observations (1/2 of the total,
the other 51 did not include 24\mum\ data) indicate that asteroid
spectral energy distributions are indeed all quite similar at these
long wavelengths, providing {\it post facto} support for the strategy.
The calibration factor we derive, which converts the instrumental
units of the 160\mum\ channel (MIPS160) to surface brightness, is
41.7~MJy/sr/MIPS160, with a formal uncertainty (uncertainty of the mean)
of 0.69~MJy/sr/MIPS160. Including the effects of the uncertainties in
the 24 and 70\mum\ calibrations, the observed repeatability of 160\mum\
measurements of a stellar calibrator and a ULIRG, and allowing for
expected uncertainties in aperture and color corrections, and modeling
uncertainties, we adopt an uncertainty of 12\% on the 160\mum\ flux
determined from an individual measurement of a source. Cross-checks
of this calibration against those of ISO measurments of ULIRGS and
nearby galaxies, and aginst IRAS measurments of asteroids, show that
the MIPS calibration is quite consistent with those earlier missions.

\acknowledgements
This work is based on observations made with the {\em Spitzer Space
Telescope}, which is operated by the Jet Propulsion Laboratory,
California Institute of Technology under NASA contract 1407. Support
for this work was provided by NASA through Contract Number \#1255094
issued by JPL/Caltech. Ephemerides were computed using the services
provided by the Solar System Dynamics group at JPL. We thank an 
anonymous reviewer for inputs which improved this paper significantly.
And, we acknowledge the wise insight of Douglas Adams, who pointed out
over 20 years ago that the answer {\em is} 42.

\begin{figure}
\epsscale{0.5}
\plotone{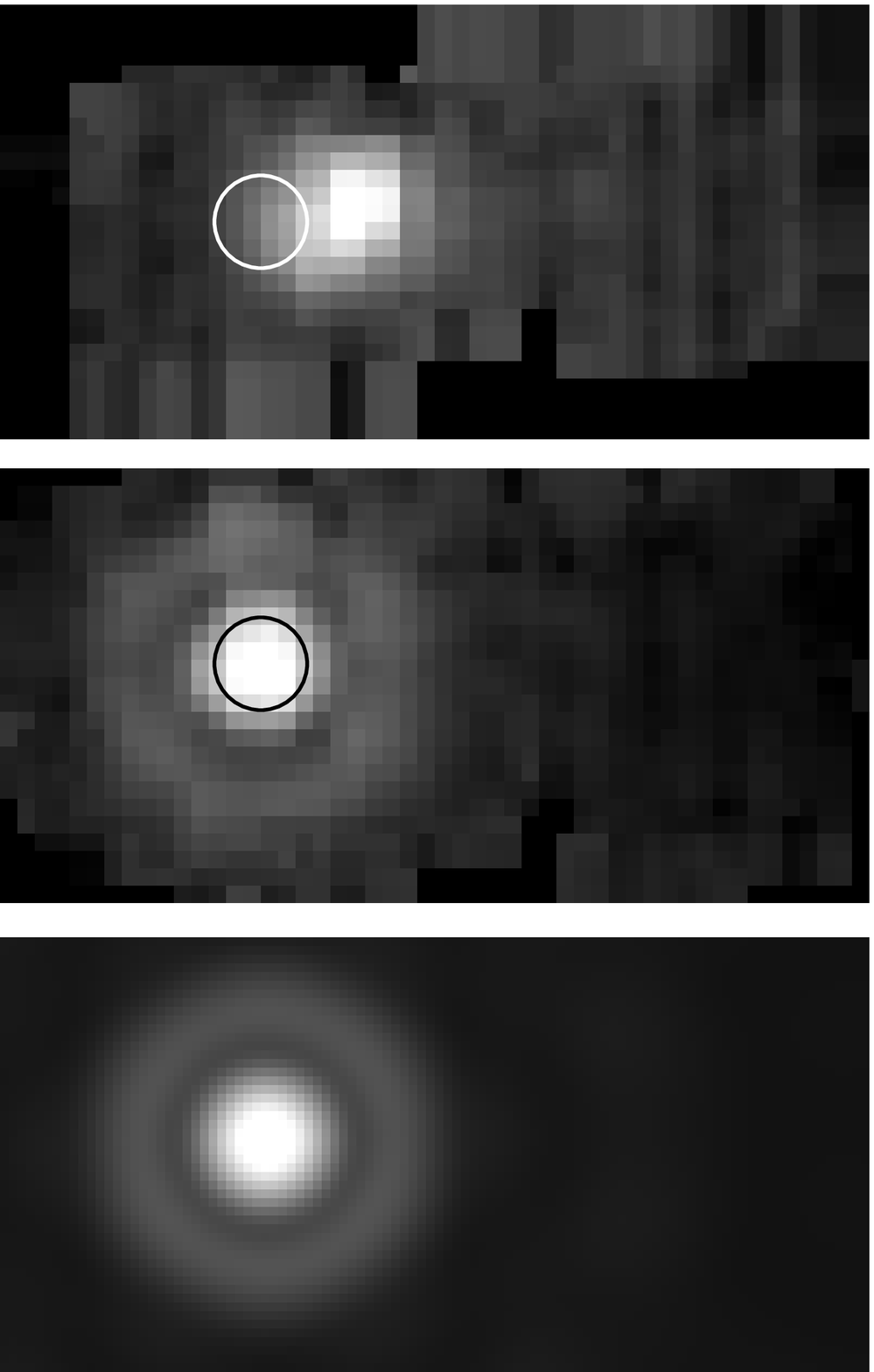}
\caption{MIPS 160\mum\ images of a star (HD 163588, top), an asteroid
(471~Papagena, middle), and an STinyTim-based model PSF (bottom).
The star image is dominated by the near-IR ghost image (see text),
while the asteroid image reveals no measureable contamination from the
ghost image. For typical asteroids, the ghost image will be $\gtrsim 2000$
times fainter, relative to the 160\mum\ image, than for stars. The
circles are centered at the pointing used in each observation. The
ghost image is always offset from the nominal pointing towards the
array centerline. The slightly different FOV of the two images (note
missing data and replicated pixels around the edge of the mosaic of the
star) results from the use of a small (3-point) map for the asteroid
observation. The mosaics were generated using a pixel scale of 8\arcsec,
$\simeq$ 1/2 the native pixel scale of the 160\mum\ array. The model
PSF was generated using STinyTim (see text) with a pixel scale of
3.2\arcsec\ and then smoothed using a boxcar 8 pixels (25.6\arcsec)
in width, equivalent to 1.6 native pixels. Each image is 6.5\arcmin\
across; the circles in the upper panels are 40\arcsec\ across.  }
\label{} \end{figure}

\begin{figure}
\epsscale{1}
\plotone{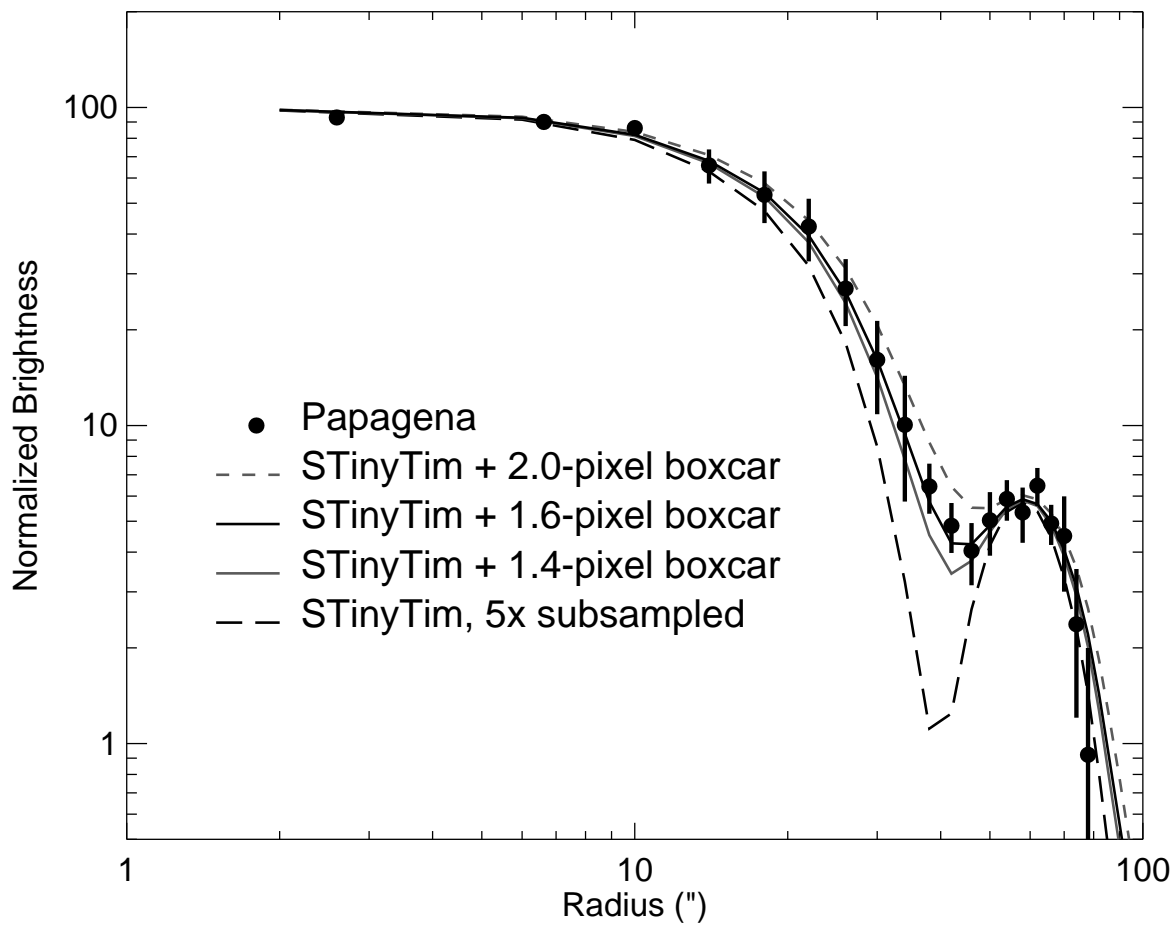}
\caption{An observed 160\mum\ PSF radial profile is compared to 4
STinyTim model PSF radial profiles. The observed profile (filled
circles) is derived from the observation of asteroid 471~Papagena shown 
in Figure~1; error bars indicate the scatter within each radial bin.
The mosaic used to generate the profile has pixels 8\arcsec\ square.
The model PSFs were generated with 3.2\farcs\ square pixels (5x
oversampled). Various smoothings were then applied to the model PSF
to match the shape of the observed PSF. Smoothing with a boxcar 
equivalent to 1.6 native pixels (25.6\farcs) results in an 
excellent match with the observed PSF.  The FWHM of the
observed PSF is 38.3\farcs, and for the model it is 38\farcs2.
}
\label{}
\end{figure}

\begin{figure}
\epsscale{1}
\plotone{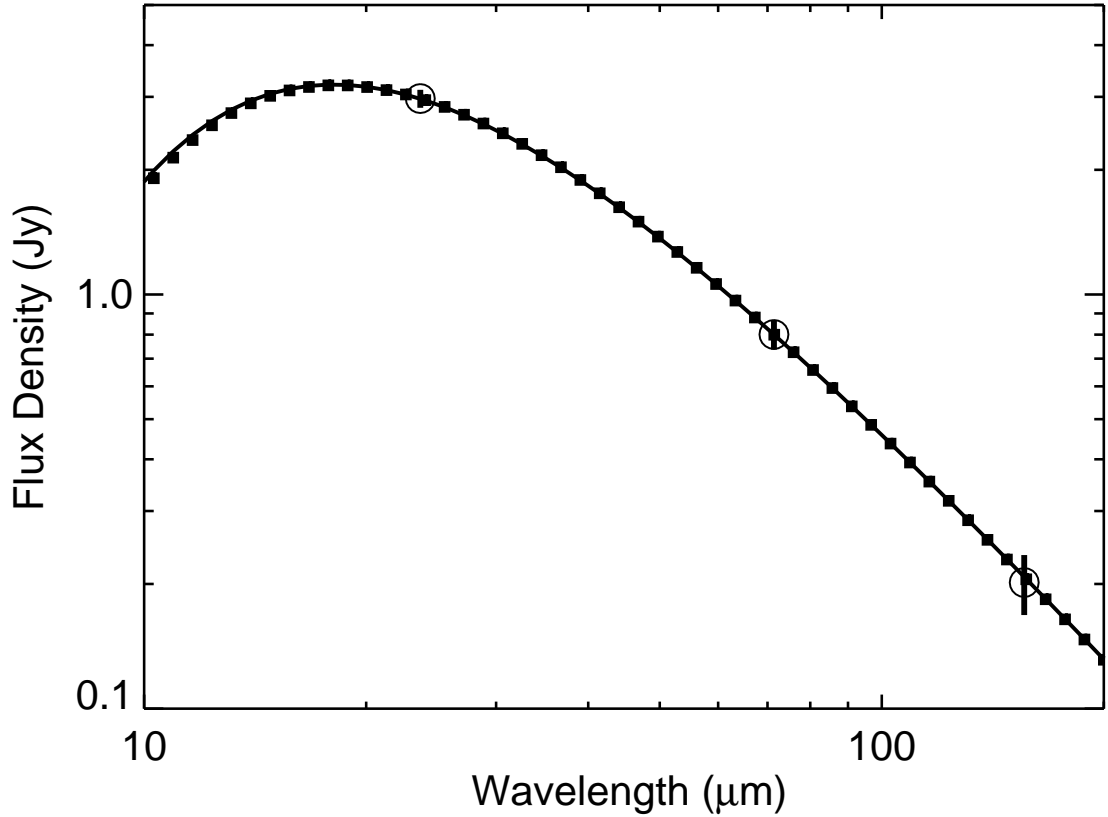}
\caption{The spectral energy distribution for asteroid 282~Clorinde is
compared to  blackbody and standard thermal model (STM) fits. The measured
SED in the MIPS channels is shown as filled circles with error bars
(the error bars are the root-sum-square of the measurement uncertainty
determined from the images and the calibration uncertainties in each
channel). The small square symbols trace a blackbody fit to the data;
the solid line shows the STM fit. The 160\mum\ point is plotted using
the calibration derived here, but was not used in the fits.
}
\label{}
\end{figure}

\begin{figure}
\epsscale{1}
\plotone{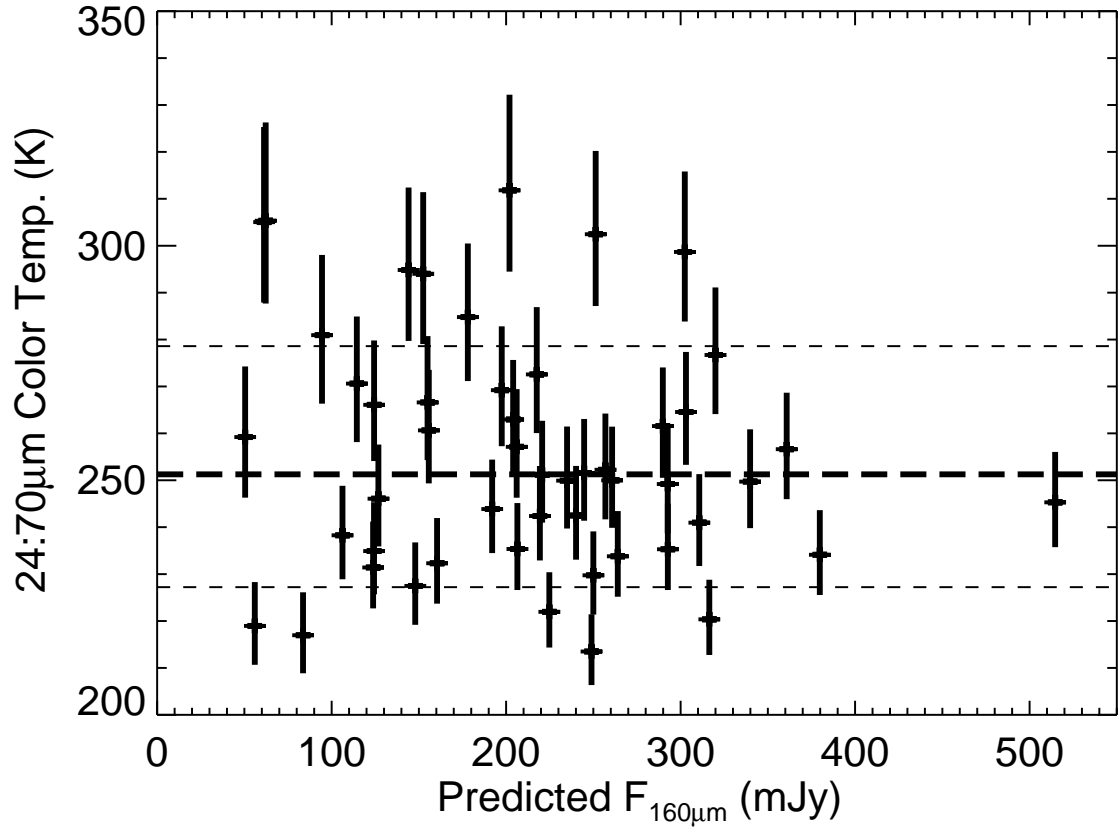}
\caption{The color temperature of those asteroids faint enough to be
observed at 24\mum. The color temperature is computed by fitting the 
24 and 70\mum\ photometry with a blackbody.  Error bars are computed 
by fitting a blackbody to the flux densities $\pm 1 \sigma$. The average 
24:70 color temperature is 251~K, and the standard deviation is 26~K (shown
by the thin dashed lines). 
}
\label{}
\end{figure}

\begin{figure}
\epsscale{1}
\plotone{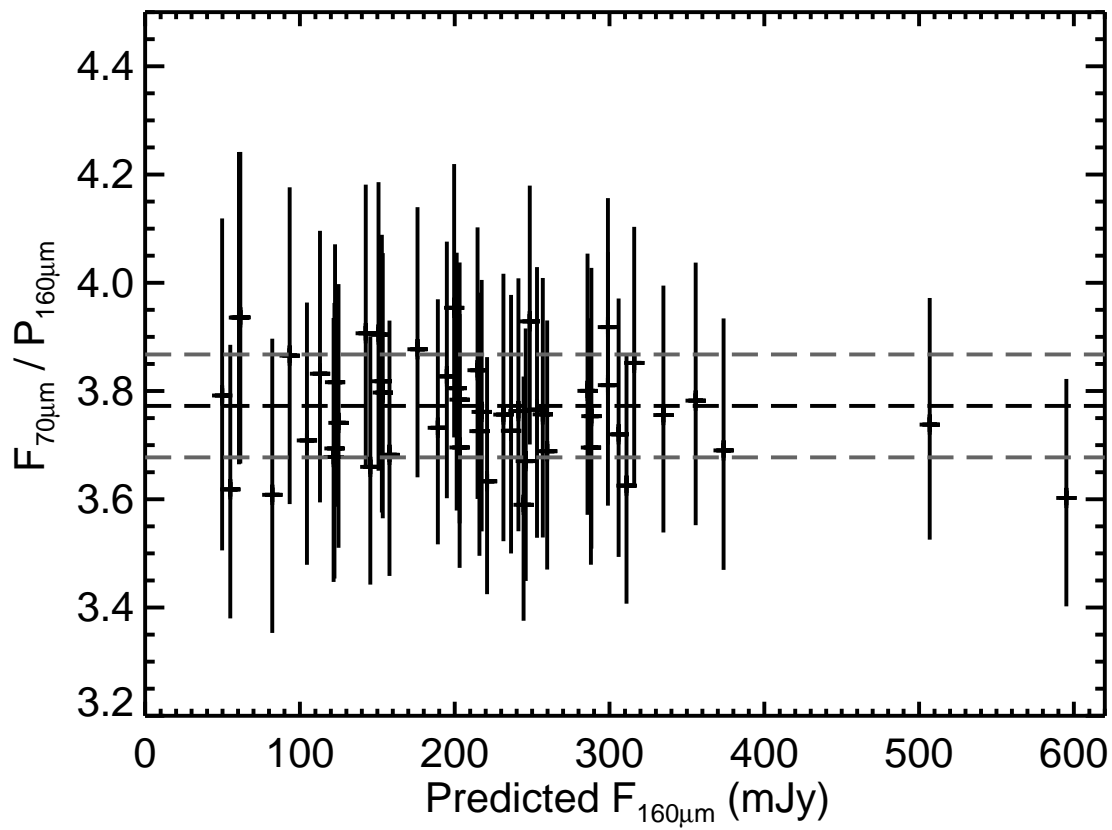}
\caption{The ratio of the measured 70\mum\ flux density to the 160\mum\ flux 
density predicted from STM fits to the 24 and 70\mum\ photometry for objects
in the 24\mum\ (faint) sample. The average 70:160\mum\ model
color (dashed line) is $3.77 \pm 0.095$, where the uncertainty is computed as the
RMS scatter of the individual predictions. The formal error on
the average color is 0.014, or about 0.4\%.
}
\label{}
\end{figure}

\begin{figure}
\epsscale{1}
\plotone{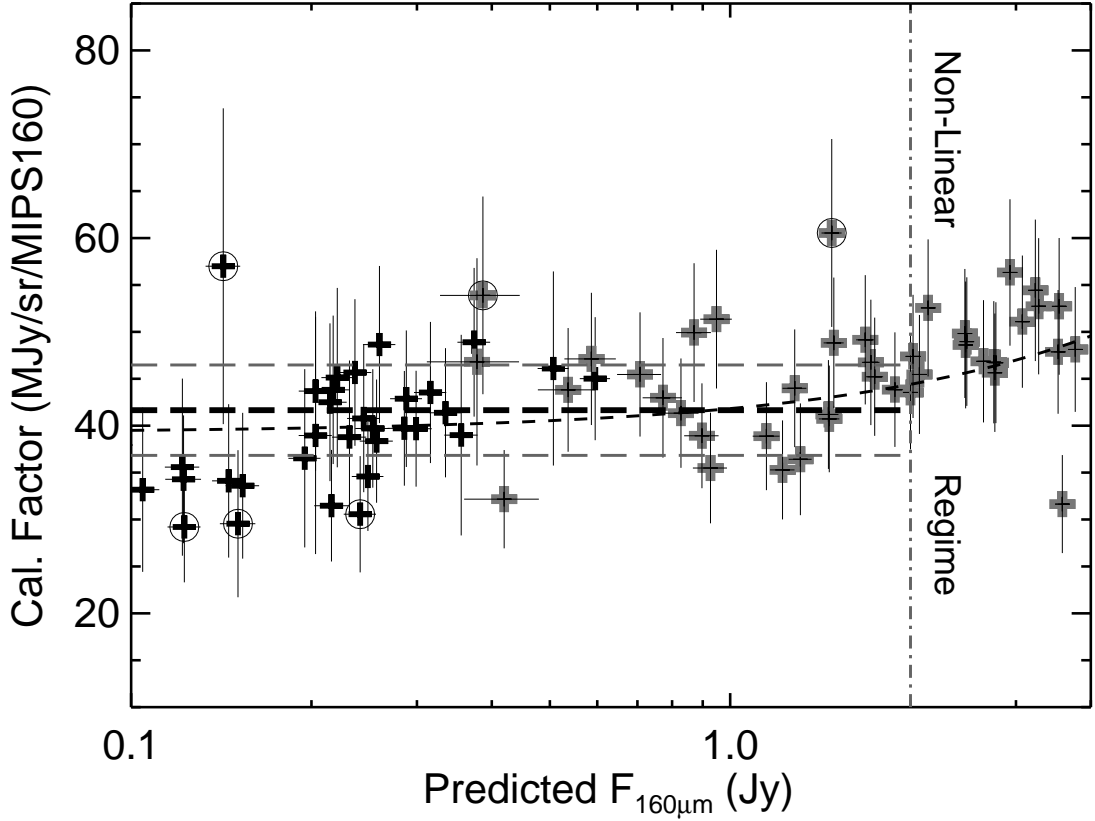}
\caption{The calibration factor for the MIPS 160\mum\ channel vs. the
predicted 160\mum\ flux density of the asteroids we observed. Black plus symbols
represent the objects in the 24\mum\ (faint) sample, which were observed
at 24, 70, and 160\mum. Grey plus symbols represent objects in the 70\mum\
(bright) sample, which was observed at 70 and 160\mum.  One-$\sigma$
uncertainties are indicated by thin error bars. Data points that are
circled were excluded from our calculation of the calibration factor
because they are discrepant at or above $1.5\sigma$. Above about 2~Jy
the response of the detectors becomes non-linear, so the points above
that are also excluded: formally, the calibration only applies below
2~Jy. The heavy long-dashed line shows the weighted-average calibration factor, $CF =
41.7\pm 0.69$~MJy/sr/MIPS160. The RMS scatter of the data is
4.82~MJy/sr/MIPS160, as shown by the thin, gray, long-dashed lines. The
short-dashed line shows a linear fit to the data (including points 
$>2$~Jy), which yields $CF = 39.2\pm 1.80$~MJy/sr/MIPS160, with a slope
of $2.58\pm0.76$~MJy/sr/MIPS160/Jy. This calibration curve can be used
to approximately calibrate targets with measured flux densities $>2$~Jy.
}
\label{}
\end{figure}

%
%
%
%
\begin{deluxetable}{lcccccccc}
\tablewidth{0pt}
\tablecaption{MIPS 160\mum\ Aperture Corrections}
\tablehead{
 & \multicolumn{6}{c}{Aperture Radius} \\ 
   \colhead{Temp.(K)} & \colhead{16\arcsec} & \colhead{24\arcsec} &
   \colhead{32\arcsec} & \colhead{40\arcsec} & \colhead{48\arcsec} &
   \colhead{64\arcsec} \\ 
}
\startdata
%
\multicolumn{4}{l}{No Sky Annulus}\\[3pt]
  10 & 4.761 &    2.657 &    2.011 &    1.776 &    1.634 &    1.402 \\ 
  30 & 4.677 &    2.610 &    1.976 &    1.745 &    1.605 &    1.355 \\ 
  50 & 4.665 &    2.603 &    1.971 &    1.740 &    1.601 &    1.348 \\ 
 150 & 4.651 &    2.595 &    1.965 &    1.735 &    1.596 &    1.341 \\ 
 250 & 4.648 &    2.593 &    1.963 &    1.734 &    1.595 &    1.340 \\ 
 500 & 4.648 &    2.593 &    1.963 &    1.734 &    1.595 &    1.339 \\ 
2000\tablenotemark{b}
     & 4.645 &    2.592 &    1.962 &    1.733 &    1.594 &    1.339 \\[3pt] 
%
\multicolumn{4}{l}{With Sky Annulus\tablenotemark{a}}\\[3pt]
  10 & 4.785 &    2.670 &    2.021 &    1.785 &    1.642 &    1.406 \\ 
  30 & 4.697 &    2.621 &    1.984 &    1.752 &    1.612 &    1.361 \\ 
  50 & 4.683 &    2.613 &    1.978 &    1.747 &    1.607 &    1.354 \\ 
 150 & 4.668 &    2.605 &    1.972 &    1.741 &    1.602 &    1.348 \\ 
 250 & 4.665 &    2.603 &    1.971 &    1.740 &    1.601 &    1.347 \\ 
 500 & 4.662 &    2.602 &    1.970 &    1.739 &    1.600 &    1.346 \\ 
2000\tablenotemark{b}
     & 4.662 &    2.602 &    1.970 &    1.739 &    1.600 &    1.345 \\ 
\enddata
\tablenotetext{a}{The sky annulus radius was 64\arcsec--128\arcsec\ for apertures up to
48\arcsec, and was 80\arcsec--160\arcsec for the 64\arcsec\ aperture.}
\tablenotetext{b}{Note that sources with near-IR:160\mum color temperatures $\ge 2000$~K are
subject to additional, large photometric uncertainty due to the contribution from the near-IR ghost image.}
\end{deluxetable}


%
%
%
\begin{deluxetable}{cccc}
\tablewidth{0pt}
\tabletypesize{\small}
\tablecaption{MIPS Color Corrections\tablenotemark{a}}
\tablehead{
 & \multicolumn{3}{c}{$\lambda_0 (\mu m$)} \\
 \cline{2-4}
 & \colhead{23.68} & \colhead{71.42} & \colhead{155.9}
}
\startdata
& \multicolumn{3}{c}{Blackbody Spectrum} \\
 T(K)\\
 \cline{1-4}
10000.0  & 1.000 &  1.000 &  1.000 \\
 1000.0  & 0.992 &  0.995 &  0.999 \\
  300.0  & 0.970 &  0.980 &  0.996 \\
  150.0  & 0.948 &  0.959 &  0.991 \\
  100.0  & 0.947 &  0.938 &  0.986 \\
   80.0  & 0.964 &  0.923 &  0.982 \\
   70.0  & 0.986 &  0.914 &  0.979 \\
   60.0  & 1.029 &  0.903 &  0.976 \\
   50.0  & 1.119 &  0.893 &  0.971 \\
   40.0  & 1.335 &  0.886 &  0.964 \\
   35.0  & 1.569 &  0.888 &  0.959 \\
   30.0  & 2.031 &  0.901 &  0.954 \\
   25.0  & 3.144 &  0.941 &  0.948 \\
   20.0  & 7.005 &  1.052 &  0.944 \\
 \cline{1-4} \\
 & \multicolumn{3}{c}{Power Law $(\nu^\beta)$} \\
 $\beta$ \\
 \cline{1-4}
   -3.0  & 0.967 &  0.933 &  0.965 \\
   -2.0  & 0.960 &  0.918 &  0.959 \\
   -1.0  & 0.961 &  0.918 &  0.959 \\
    0.0  & 0.967 &  0.932 &  0.965 \\
    1.0  & 0.981 &  0.959 &  0.979 \\
    2.0  & 1.001 &  1.001 &  1.000 \\
    3.0  & 1.027 &  1.057 &  1.029 \\
\enddata
\tablenotetext{a}{Divide measured fluxes by these
values to compute the corrected monochromatic flux density.}
\end{deluxetable}


%
%
%
%
\begin{deluxetable}{rlcccccccccccc}
\tabletypesize{\scriptsize} 
\rotate
\tablewidth{0pt}
\tablecaption{24\mum\ (Faint) Sample}
\tablehead{
   \colhead{} & \colhead{} & \colhead{} & \multicolumn{2}{c}{AORKEYS\tablenotemark{a}} \\
   \colhead{\#} & \colhead{Asteroid} & \colhead{Obs. Date} & \colhead{24+70} & \colhead{160} &
   \colhead{$F_{24}$\tablenotemark{b}}  &  \colhead{$err_{24}$\tablenotemark{b}} & \colhead{$F_{70}$\tablenotemark{b}} & 
   \colhead{$err_{70}$\tablenotemark{b}} & \colhead{$P_{160}$\tablenotemark{b}}  & \colhead{$err_{P160}$\tablenotemark{b}} &  
   \colhead{MIPS160\tablenotemark{c}} & \colhead{$err_{M160}$\tablenotemark{c}} & \colhead{CF\tablenotemark{d}} \\ [-12pt]
}
\startdata
%
 186 & Celuta        &  2004-02-23 &          &  9064960 &  4.612 &  0.231 &  1.345 &  0.139 &  0.356 &  0.049 &  1.516 &  0.406 & 38.99 \\
 248 & Lameia        &  2004-02-23 &          &  9065216 &  3.585 &  0.179 &  1.082 &  0.115 &  0.288 &  0.041 &  1.516 &  0.491 & 31.60 \\
 443 & Photographica &  2004-02-23 &          &  9065728 &  2.089 &  0.104 &  0.584 &  0.063 &  0.153 &  0.022 &  0.628 &  0.390 & 40.46 \\
 186 & Celuta        &  2004-03-18 &          &  9193216 &  2.700 &  0.135 &  0.902 &  0.093 &  0.246 &  0.034 &  1.053 &  0.377 & 38.82 \\
  25 & Phocaea       &  2004-03-18 &          &  9193728 &  3.618 &  0.181 &  1.138 &  0.117 &  0.306 &  0.042 &  1.961 &  0.869 & 25.93 \\
 432 & Pythia        &  2004-03-18 &          &  9193984 &  3.186 &  0.159 &  1.127 &  0.116 &  0.311 &  0.042 &  0.827 &  0.372 & 62.50 \\
 284 & Amalia        &  2004-04-07 &          &  9460224 &  3.805 &  0.190 &  1.086 &  0.112 &  0.286 &  0.039 &  1.196 &  0.249 & 39.72 \\
 783 & Nora          &  2004-04-07 &          &  9460736 &  6.159 &  0.308 &  1.895 &  0.195 &  0.507 &  0.069 &  1.827 &  0.445 & 46.09 \\
 432 & Pythia        &  2004-04-07 &          &  9460992 &  2.329 &  0.116 &  0.589 &  0.062 &  0.151 &  0.021 &  0.848 &  0.224 & 29.55 \\
1584 & Fuji          &  2004-04-08 &          &  9460480 &  1.574 &  0.079 &  0.433 &  0.046 &  0.113 &  0.016 &  0.718 &  0.223 & 26.16 \\
1584 & Fuji          &  2004-05-02 &          &  9664512 &  0.993 &  0.050 &  0.242 &  0.027 &  0.061 &  0.009 &  0.577 &  0.159 & 17.71 \\
  60 & Echo          &  2004-05-07 &          &  9665792 &  2.890 &  0.145 &  0.869 &  0.091 &  0.231 &  0.032 &  0.992 &  0.234 & 38.78 \\
1137 & Raissa        &  2004-05-07 &          &  9666048 &  0.654 &  0.033 &  0.189 &  0.022 &  0.050 &  0.008 &  0.328 &  0.114 & 25.22 \\
1584 & Fuji          &  2004-06-02 &          &  9810176 &  0.557 &  0.028 &  0.199 &  0.022 &  0.055 &  0.008 &  0.309 &  0.110 & 29.62 \\
 453 & Tea           &  2004-06-02 &          &  9809920 &  1.572 &  0.079 &  0.532 &  0.055 &  0.145 &  0.020 &  0.709 &  0.178 & 34.12 \\
 113 & Amalthea      &  2004-06-04 &          &  9810432 &  2.371 &  0.119 &  0.878 &  0.091 &  0.244 &  0.034 &  0.997 &  0.227 & 40.77 \\
 623 & Chimaera      &  2004-06-18 &          &  9935104 &  2.318 &  0.116 &  0.751 &  0.077 &  0.203 &  0.028 &  0.773 &  0.176 & 43.69 \\
 572 & Rebekka       &  2004-06-19 &          &  9935360 &  0.820 &  0.041 &  0.296 &  0.033 &  0.082 &  0.012 &  0.337 &  0.115 & 40.55 \\
 273 & Atropos       &  2004-06-22 &          &  9934848 &  1.672 &  0.084 &  0.468 &  0.049 &  0.123 &  0.017 &  0.698 &  0.162 & 29.20 \\
 623 & Chimaera      &  2004-07-09 &          & 10085120 &  1.353 &  0.068 &  0.448 &  0.047 &  0.122 &  0.017 &  0.569 &  0.150 & 35.58 \\
 138 & Tolosa        &  2004-07-09 &          & 10084864 &  5.888 &  0.294 &  2.144 &  0.218 &  0.595 &  0.080 &  2.197 &  0.449 & 45.01 \\
 234 & Barbara       &  2004-07-29 &          & 11779328 &  2.735 &  0.137 &  0.818 &  0.085 &  0.217 &  0.030 &  0.825 &  0.183 & 43.82 \\
 376 & Geometria     &  2004-08-23 &          & 11896576 &  2.291 &  0.115 &  0.803 &  0.082 &  0.221 &  0.030 &  0.814 &  0.193 & 45.12 \\
 376 & Geometria     &  2004-08-24 &          & 11896832 &  2.278 &  0.114 &  0.706 &  0.072 &  0.189 &  0.026 &  0.483 &  0.181 & 65.06 \\
 364 & Isara         &  2004-09-21 &          & 12058624 &  3.301 &  0.165 &  0.789 &  0.084 &  0.200 &  0.028 &  0.541 &  0.236 & 61.34 \\
 189 & Phthia        &  2004-09-21 &          & 12058112 &  2.699 &  0.135 &  0.766 &  0.079 &  0.201 &  0.028 &  0.512 &  0.213 & 65.34 \\
 856 & Backlunda     &  2004-10-14 &          & 12428544 &  2.642 &  0.132 &  0.769 &  0.080 &  0.203 &  0.028 &  0.867 &  0.254 & 38.95 \\
 364 & Isara         &  2004-10-14 &          & 12232448 &  4.706 &  0.235 &  1.171 &  0.121 &  0.299 &  0.041 &  0.766 &  0.322 & 64.89 \\
1137 & Raissa        &  2004-10-14 &          & 12232960 &  1.215 &  0.061 &  0.388 &  0.041 &  0.105 &  0.015 &  0.524 &  0.138 & 33.18 \\
  60 & Echo          &  2004-11-04 &          & 12393728 &  3.204 &  0.160 &  0.953 &  0.100 &  0.253 &  0.035 &  1.060 &  0.223 & 39.70 \\
  60 & Echo-1        &  2004-11-04 &          & 12544000 &  3.285 &  0.164 &  1.064 &  0.109 &  0.288 &  0.039 &  1.116 &  0.241 & 42.88 \\
  60 & Echo-2        &  2004-11-04 &          & 12544512 &  3.209 &  0.160 &  0.965 &  0.100 &  0.257 &  0.035 &  1.113 &  0.241 & 38.35 \\
 189 & Phthia        &  2004-11-05 &          & 12393984 &  2.035 &  0.102 &  0.583 &  0.061 &  0.154 &  0.021 &  0.759 &  0.187 & 33.60 \\
 131 & Vala          &  2004-11-29 &          & 12870656 &  1.364 &  0.068 &  0.361 &  0.040 &  0.093 &  0.014 &  0.409 &  0.114 & 37.94 \\
 198 & Ampella       &  2004-12-02 &          & 12870144 &  2.823 &  0.141 &  0.881 &  0.091 &  0.236 &  0.032 &  0.860 &  0.186 & 45.67 \\
 198 & Ampella       &  2005-01-02 &          & 13070336 &  4.228 &  0.211 &  1.379 &  0.141 &  0.374 &  0.051 &  1.270 &  0.270 & 48.89 \\
 470 & Kilia         &  2005-01-02 &          & 13070848 &  1.764 &  0.088 &  0.581 &  0.060 &  0.158 &  0.022 &  0.593 &  0.203 & 44.20 \\
 248 & Lameia        &  2005-01-02 &          & 13070592 &  3.040 &  0.152 &  0.907 &  0.094 &  0.241 &  0.033 &  1.311 &  0.305 & 30.56 \\
 376 & Geometria     &  2005-01-24 & 13107456 & 13107200 &  1.391 &  0.070 &  0.452 &  0.048 &  0.122 &  0.017 &  0.593 &  0.137 & 34.29 \\
 556 & Phyllis       &  2005-01-24 & 13107968 & 13107712 &  2.208 &  0.110 &  0.557 &  0.059 &  0.142 &  0.020 &  0.416 &  0.116 & 56.99 \\
 757 & Portlandia    &  2005-01-29 & 13108480 & 13108224 &  2.932 &  0.147 &  0.958 &  0.099 &  0.260 &  0.035 &  0.887 &  0.193 & 48.65 \\
 443 & Photographica &  2005-03-01 & 13307648 & 13307392 &  1.525 &  0.076 &  0.467 &  0.050 &  0.125 &  0.018 &  0.241 &  0.212 & 86.22 \\
 495 & Eulalia       &  2005-03-02 & 13308160 & 13307904 &  2.697 &  0.135 &  0.746 &  0.077 &  0.195 &  0.027 &  0.887 &  0.232 & 36.53 \\
 512 & Taurinensis   &  2005-03-02 & 13307136 & 13306880 &  0.977 &  0.049 &  0.238 &  0.026 &  0.061 &  0.009 &  0.204 &  0.445 & 49.38 \\
 443 & Photographica &  2005-04-05 & 13443840 & 13443584 &  2.615 &  0.131 &  0.682 &  0.071 &  0.176 &  0.024 &  0.872 &  0.315 & 33.55 \\
 118 & Peitho        &  2005-05-14 & 13637120 & 13636864 &  4.531 &  0.227 &  1.217 &  0.126 &  0.316 &  0.043 &  1.206 &  0.262 & 43.54 \\
 584 & Semiramis     &  2005-05-14 & 13636608 & 13636352 &  2.578 &  0.129 &  0.805 &  0.084 &  0.216 &  0.030 &  1.141 &  0.257 & 31.47 \\
 435 & Ella          &  2005-05-15 & 13637632 & 13637376 &  4.176 &  0.209 &  1.257 &  0.129 &  0.335 &  0.046 &  1.345 &  0.288 & 41.38 \\
 282 & Clorinde      &  2005-06-18 & 15244800 & 15244544 &  3.021 &  0.151 &  0.824 &  0.087 &  0.215 &  0.030 &  0.840 &  0.193 & 42.50 \\
 126 & Velleda       &  2005-06-18 & 15245824 & 15245568 &  3.969 &  0.198 &  0.976 &  0.101 &  0.248 &  0.034 &  1.194 &  0.257 & 34.59 \\
 877 & Walkure       &  2005-06-18 & 15245312 & 15245056 &  4.043 &  0.202 &  1.139 &  0.117 &  0.299 &  0.041 &  1.252 &  0.262 & 39.68 \\
\enddata
\tablenotetext{a}{Inique identifier for data in the Spitzer archive. Where only the 160 AORKEY is given, the same key 
     applies to the 24 and 70\mum\ data.}
\tablenotetext{b}{Color-corrected flux densities and uncertainties in Jy. The uncertainties include the uncertainty in the 
absolute calibration of the 24 and 70\mum\ bands (2\% and 5\%, respectively).}
\tablenotetext{c}{Color-corrected 160\mum\ channel flux density and uncertainty, in instrumental units.}
\tablenotetext{d}{Calibration factor derived from each observation, MJy/sr/MIPS160.}
\end{deluxetable}


%
%
%
%
\begin{deluxetable}{rlcccccccccc}
\tablewidth{0pt}
\tabletypesize{\scriptsize} 
\tablecaption{70\mum\ (Bright) Sample}
\tablehead{
   \colhead{} & \colhead{} & \colhead{} & \multicolumn{2}{c}{AORKEYS\tablenotemark{a}} \\ [2pt]
   \colhead{\#} & \colhead{Asteroid} & \colhead{Obs. Date} & \colhead{70} & \colhead{160} &
   \colhead{$F_{70}$\tablenotemark{b}} & \colhead{$err_{70}$\tablenotemark{b}} & 
   \colhead{$P_{160}$\tablenotemark{b}}  & \colhead{$err_{P160}$\tablenotemark{b}} &
   \colhead{MIPS160\tablenotemark{c}} & \colhead{$err_{M160}$\tablenotemark{c}} & \colhead{CF\tablenotemark{d}} \\ [-10pt]
}
\startdata
%
 337 & Devosa        &  2003-12-13 &          &  8780288 &  5.573 &  0.285 &  1.476 &  0.057 &  4.056 &  0.307 & 60.49 \\
1584 & Fuji          &  2003-12-13 &          &  8779520 &  0.478 &  0.025 &  0.127 &  0.057 &  0.603 &  0.063 & 34.90 \\
 752 & Sulamitis     &  2003-12-13 &          &  8780032 &  0.654 &  0.040 &  0.173 &  0.066 &  0.624 &  0.074 & 46.14 \\
 198 & Ampella       &  2004-01-25 &          &  8811776 &  2.667 &  0.144 &  0.706 &  0.060 &  2.585 &  0.108 & 45.43 \\
  83 & Beatrix       &  2004-01-25 &          &  8812032 &  3.384 &  0.182 &  0.896 &  0.059 &  3.830 &  0.142 & 38.90 \\
 345 & Tercidina     &  2004-01-25 &          &  8812288 &  4.336 &  0.233 &  1.149 &  0.059 &  4.913 &  0.232 & 38.85 \\
  25 & Phocaea       &  2004-02-23 &          &  9065472 &  1.425 &  0.087 &  0.378 &  0.066 &  1.342 &  0.199 & 46.77 \\
 345 & Tercidina     &  2004-02-23 &          &  9065984 &  6.478 &  0.338 &  1.716 &  0.058 &  6.106 &  0.216 & 46.71 \\
 783 & Nora          &  2004-03-18 &          &  9194240 &  0.996 &  0.056 &  0.264 &  0.062 &  1.641 &  0.441 & 26.72 \\
  60 & Echo          &  2004-06-01 &          &  9809408 &  1.458 &  0.078 &  0.386 &  0.059 &  1.192 &  0.132 & 53.85 \\
  18 & Melpomene     &  2004-06-18 &          &  9934592 &  5.618 &  0.290 &  1.488 &  0.057 &  5.069 &  0.173 & 48.79 \\
   7 & Iris          &  2004-06-20 &          &  9934080 & 14.209 &  0.732 &  3.764 &  0.057 & 13.010 &  0.244 & 48.08 \\
 505 & Cava          &  2004-07-11 &          & 10084608 &  4.834 &  0.258 &  1.280 &  0.059 &  4.842 &  0.167 & 43.95 \\
  40 & Harmonia      &  2004-07-11 &          & 10084352 &  7.618 &  0.396 &  2.018 &  0.058 &  7.084 &  0.143 & 47.35 \\
  40 & Harmonia      &  2004-07-29 &          & 11779840 & 10.392 &  0.536 &  2.753 &  0.057 &  9.792 &  0.225 & 46.72 \\
  20 & Massalia      &  2004-07-29 &          & 11778816 &  6.575 &  0.338 &  1.742 &  0.057 &  6.406 &  0.153 & 45.18 \\
  40 & Harmonia      &  2004-08-23 &          & 11896064 & 13.304 &  0.683 &  3.524 &  0.057 & 12.246 &  0.202 & 47.83 \\
  20 & Massalia      &  2004-08-23 &          & 11895552 &  5.503 &  0.284 &  1.458 &  0.057 &  5.885 &  0.171 & 41.16 \\
  19 & Fortuna       &  2004-09-15 &          & 12057600 & 12.648 &  0.646 &  3.351 &  0.057 & 20.179 &  0.677 & 27.60 \\
  12 & Victoria      &  2004-09-22 &          & 12057088 &  3.295 &  0.176 &  0.873 &  0.059 &  7.575 &  2.197 & 19.15 \\
   3 & Juno          &  2004-09-26 &          & 12059648 & 21.511 &  1.107 &  5.698 &  0.057 & 33.718 &  2.360 & 28.09 \\
  12 & Victoria      &  2004-10-14 &          & 12231936 &  4.941 &  0.256 &  1.309 &  0.058 &  5.977 &  0.437 & 36.39 \\
 313 & Chaldaea      &  2004-11-04 &          & 12393216 &  1.582 &  0.086 &  0.419 &  0.060 &  2.167 &  0.156 & 32.14 \\
  12 & Victoria      &  2004-11-04 &          & 12392704 &  5.369 &  0.275 &  1.422 &  0.057 &  9.404 &  0.938 & 25.14 \\
 433 & Eros          &  2004-11-29 &          & 12869120 &  2.212 &  0.115 &  0.586 &  0.058 &  2.069 &  0.104 & 47.09 \\
  83 & Beatrix       &  2005-01-02 &          & 13071360 &  6.339 &  0.328 &  1.679 &  0.058 &  5.682 &  0.152 & 49.12 \\
 433 & Eros          &  2005-01-02 &          & 13071104 &  3.578 &  0.185 &  0.948 &  0.058 &  3.070 &  0.116 & 51.31 \\
  21 & Lutetia       &  2005-01-24 & 13106944 & 13106688 &  7.806 &  0.401 &  2.068 &  0.057 &  7.566 &  0.184 & 45.42 \\
  12 & Victoria      &  2005-03-02 & 13306624 & 13306368 &  3.497 &  0.193 &  0.926 &  0.061 &  4.342 &  0.334 & 35.46 \\
   7 & Iris          &  2005-04-12 & 13442304 & 13442048 & 13.522 &  0.700 &  3.582 &  0.058 & 18.829 &  1.420 & 31.62 \\
  42 & Isis          &  2005-05-16 & 13636096 & 13635840 & 11.057 &  0.567 &  2.929 &  0.057 &  8.648 &  0.174 & 56.29 \\
   6 & Hebe          &  2005-06-18 & 15244288 & 15244032 & 13.356 &  0.685 &  3.538 &  0.057 & 11.163 &  0.208 & 52.67 \\
 471 & Papagena\tablenotemark{e}
                     &  2005-07-27 & 15418112 & 15417856 & 10.433 &  0.535 &  2.764 &  0.057 & 10.070 &  0.188 & 45.61 \\
 471 & Papagena\tablenotemark{e}
                     &  2005-07-27 & 15418624 & 15418368 & 12.347 &  0.634 &  3.271 &  0.057 & 10.317 &  0.163 & 52.69 \\
 471 & Papagena\tablenotemark{e}
                     &  2005-07-27 & 15419136 & 15418880 & 11.598 &  0.595 &  3.072 &  0.057 & 10.002 &  0.171 & 51.05 \\
  23 & Thalia\tablenotemark{e}
                     &  2005-07-28 & 15419648 & 15419392 &  2.914 &  0.152 &  0.772 &  0.058 &  2.987 &  0.143 & 42.94 \\
  23 & Thalia\tablenotemark{e}
                     &  2005-07-28 & 15420160 & 15419904 &  3.283 &  0.172 &  0.870 &  0.058 &  2.897 &  0.138 & 49.88 \\
  23 & Thalia\tablenotemark{e}
                     &  2005-07-28 & 15420672 & 15420416 &  3.122 &  0.163 &  0.827 &  0.058 &  3.327 &  0.099 & 41.30 \\
 313 & Chaldaea      &  2005-08-26 & 15813632 & 15813376 &  4.613 &  0.238 &  1.222 &  0.057 &  5.761 &  0.290 & 35.26 \\
  41 & Daphne        &  2005-08-27 & 15813120 & 15812864 &  7.544 &  0.388 &  1.999 &  0.057 &  7.634 &  0.184 & 43.51 \\
 138 & Tolosa        &  2005-08-29 & 15814656 & 15814400 &  2.023 &  0.108 &  0.536 &  0.059 &  2.035 &  0.106 & 43.78 \\
 433 & Eros          &  2005-09-04 & 15814144 & 15813888 &  0.509 &  0.032 &  0.135 &  0.068 &  0.968 &  0.308 & 23.18 \\
  42 & Isis\tablenotemark{e}
                     &  2005-11-09 & 16259584 & 16258816 & 10.474 &  0.537 &  2.775 &  0.057 &  9.879 &  0.168 & 46.68 \\
  42 & Isis\tablenotemark{e}
                     &  2005-11-09 & 16259840 & 16259072 & 12.191 &  0.624 &  3.229 &  0.057 &  9.869 &  0.199 & 54.38 \\
  42 & Isis\tablenotemark{e}
                     &  2005-11-09 & 16260096 & 16259328 &  9.367 &  0.480 &  2.481 &  0.057 &  8.428 &  0.218 & 48.93 \\
  20 & Massalia\tablenotemark{e}
                     &  2005-11-30 & 16465408 & 16464384 &  9.298 &  0.478 &  2.463 &  0.057 &  8.222 &  0.144 & 49.78 \\
  20 & Massalia\tablenotemark{e}
                     &  2005-11-30 & 16465664 & 16464640 &  8.073 &  0.415 &  2.138 &  0.057 &  6.768 &  0.148 & 52.51 \\
  20 & Massalia\tablenotemark{e}
                     &  2005-11-30 & 16465920 & 16464896 &  9.324 &  0.478 &  2.470 &  0.057 &  8.452 &  0.188 & 48.57 \\
  20 & Massalia\tablenotemark{e}
                     &  2005-11-30 & 16466176 & 16465152 &  7.104 &  0.366 &  1.882 &  0.057 &  7.140 &  0.166 & 43.81 \\
  85 & Io            &  2006-01-11 & 16617984 & 16617728 &  5.520 &  0.286 &  1.462 &  0.058 &  5.973 &  0.148 & 40.69 \\
  51 & Nemausa       &  2006-01-12 & 16618496 & 16618240 &  9.987 &  0.511 &  2.645 &  0.057 &  9.389 &  0.210 & 46.83 \\
\enddata
\tablenotetext{a}{Inique identifier for data in the Spitzer archive. Where only the 160 AORKEY is given, the same key
     applies to the 70\mum\ data.}
\tablenotetext{b}{Color-corrected flux densities and uncertainties in Jy. The uncertainties include the uncertainty in 
     the absolute calibration of the 24 and 70\mum\ bands (2\% and 5\%, respectively).}
\tablenotetext{c}{Color-corrected 160\mum\ channel flux density and uncertainty, in instrumental units.}
\tablenotetext{d}{Calibration factor derived from each observation, MJy/sr/MIPS160.}
\tablenotetext{e}{These objects were observed several times on the given date. The Papagena and Thalia observations
     were taken without interruption; those for Isis and Massalia were spaced by about 2 hours. Lightcurve variations
     caused by the shape of these targets are predicted to contribute about 5\% to the observed variation except for 
     all except Papagena, where the lightcurve should have only contributed about a 1\% variation over the observing
     interval.}
\end{deluxetable}



\clearpage
\begin{thebibliography}{}
\bibitem[]{} Beichmann, C.A. et al. 1985.  Infrared Astronomical Satellite
        (IRAS) Catalogs and Atlases Explanatory Supplement", ed. C. A. Beichman,
        G. Neugebauer, H. J. Habing, P. E. Clegg, and T. J. Chester.
        U. S. Government Printing Office.
\bibitem[]{} Bowell, E. et al. 1989, in  Asteroids II, ed. R. P.Binzel, 
        T. Gehrels, T., \& M. S. Matthews (Tucson: Univ. Arizona Press)
\bibitem[]{} Campins, H. et al. 1994, AJ 108, 2318
\bibitem[]{} Dale, D.A. et al. 2005, ApJ 633, 857
\bibitem[]{} Dole, H., et al. 2006, A\&A 451, 417
\bibitem[]{} Dale, D.A. et al. 2007, ApJ 655, 863
\bibitem[]{} Diolaiti, E. et al. 2000, A\&AS 147, 335
\bibitem[]{} Engelbracht, C.W. et al. 2007, this issue
\bibitem[]{} Fazio, G. et al. 2004 ApJS 154, 10
\bibitem[]{} Fernandez, Y.F. et al. 2002, AJ 123, 2050
\bibitem[]{} Fixsen, D.J. et al. 1997, ApJ 490, 482
\bibitem[]{} Gordon, K.D., et al. 2007, in prep.
\bibitem[]{} Gordon, K.D., et al. 2007, this issue.
\bibitem[]{} Gordon, K.D. et al.  2005, \pasp\ 117, 503
\bibitem[]{} Gordon, K.D., et al. 2006, ApJ 638, 87
\bibitem[]{} Haas, M. et al, 1998, A\&A 338, L33
\bibitem[]{} Harris, A.W. 1998. Icarus 131, 291
\bibitem[]{} Hauser, M.G. et al., 1998 ApJ 508, 25-43)
\bibitem[]{} Hinz, J.L. et al 2004, ApJS, 154, 259
\bibitem[]{} Hippelein, H. et al., 2003, A\&A 315, L82
\bibitem[]{} Hora, J.L. et al. 2004, SPIE 5487, 77
\bibitem[]{} Houck, J.R. et al. 2004, ApJS 154, 18
\bibitem[]{} Klaas, U. et al. 2001, A\&A 379, 823
\bibitem[]{} Krist, J. 2002, Tiny Time/SIRTF User's Guide (Pasadena: SSC)
\bibitem[]{} Lebofsky, L. A, et al. 1986, Icarus 68, 239.
\bibitem[]{} Lebofsky, L. A, \& Spencer, J. R. 1989,
        in Asteroids II, ed. R. P.Binzel, T. Gehrels, T., \& M. S. Matthews
        (Tucson: Univ. Arizona Press)
\bibitem[]{} Lebofsky, L.A. et al. 1986, Icarus 68, 1694
\bibitem[]{} Lumme, K. and E. Bowell 1981, AJ 86, 1694
\bibitem[]{} Mather, J.C. et al. 1999, ApJ 512, 511
\bibitem[]{} M\"uller, T.G, and J.S.V. Lagerros, 1998, A\&A 338, 340
\bibitem[]{} M\"uller, T.G, and J.S.V. Lagerros, 2002, A\&A 381, 324
\bibitem[]{} Neugebauer, G. et al. 1984, ApJ 278, 1
\bibitem[]{} Reach, W.T. et al. 2005, PASP 117, 978
\bibitem[]{} Rieke, G.H. et al. 2004, ApJS 154, 25
\bibitem[]{} Rieke, G.H. et al. 2007, in prep.
\bibitem[]{} Schulz, B. et al. 2002, A\&A 381, 1110
\bibitem[]{} Stansberry, J.A. et al. 2006, ApJ ~643, 556
\bibitem[]{} Stickel, M. et al. 2004, A\&A 422, 39
\bibitem[]{} Tedesco, E. F. et al. 2002, \aj ~123, 2056
\bibitem[]{} Werner, M.W. et al. 2004, ApJS 154, 1

\end{thebibliography}
\end{document}